%% file: ms_Schumann_final.tex
\documentclass[twocolumn,aps,prd,preprintnumbers,superscriptaddress,nofootinbib]{revtex4-1}
\usepackage{aas_macros}

\usepackage{hyperref}
\usepackage{amsmath,amssymb,amsfonts,mathrsfs}
\usepackage[dvipdfmx]{graphicx}
\usepackage{color}
\usepackage{enumerate} 
\usepackage{bm}
\usepackage{array}
\usepackage{siunitx}
\usepackage{float}

\newcommand{\bfOmega}{\mbox{\boldmath$\Omega$}}

\begin{document}
\title{Correlated magnetic noise from anisotropic lightning sources and the detection of stochastic gravitational waves}
\author{Yoshiaki Himemoto}
\affiliation{Department of Liberal Arts and Basic Sciences, College of Industrial Technology, Nihon University, Narashino, Chiba 275-8576, Japan}
\author{Atsushi Taruya}
\affiliation{Center for Gravitational Physics, Yukawa Institute for Theoretical Physics, Kyoto University, Kyoto 606-8502, Japan}
\affiliation{
Kavli Institute for the Physics and Mathematics of the Universe, Todai Institutes for Advanced Study, the University of Tokyo, Kashiwa, Chiba 277-8583, Japan (Kavli IPMU, WPI)}
%
%
%
%
\date{\today}
\begin{abstract} 
Direct detection of gravitational waves (GWs) from compact binary systems suggests that the merger rate of such events is large, and the sum of their GWs can be viewed as stochastic signals.
Because of its random nature, cross-correlating the signals from multiple detectors is essential to disentangle the GWs from instrumental noise. However, the global magnetic fields in the Earth-ionosphere cavity produce the environmental disturbances at low-frequency bands, known as Schumann resonances, and coupled with GW detectors, they  potentially contaminate the stochastic GW signal as a correlated noise. Previously, we have presented a simple analytical model to estimate its impact on the detection of stochastic GWs. Here, extending the analysis to further take account of the effects of anisotropic lightning source distributions, we present a comprehensive study of the impact of correlated magnetic noise at low-frequency bands, including non-tensor-type GWs, as well as circularly polarized tensor-type GWs. We find that as opposed to a naive expectation, the impact of correlated magnetic noise does not always increase with anisotropies in the lighting source distribution. Even in the presence of large anisotropies, there is a robust detector pair for which the amplitude of correlated magnetic noise becomes comparable to or well below detectable amplitude of stochastic GWs. The results indicate that the properties of the correlated magnetic noise depend crucially on both the geometrical and geographical setup of the detector's pair, and Virgo and KAGRA would be potentially the most insensitive detector pair against the correlated magnetic for both tensor- and non-tensor-type stochastic GWs. 

\end{abstract}

\preprint{YITP-19-77}
\maketitle


\section{Introduction}
\label{sec:intro}
%
%
Since the first discovery of a gravitational wave (GW) event by laser interferometer LIGO (Hanford and Livingston, \cite{2010CQGra..27h4006H, 2015CQGra..32g4001L}), there is a growing interest in detecting the stochastic gravitational wave backgrounds. Stochastic GWs are basically produced by an incoherent superposition of an extremely large number of GWs from the unresolved astrophysical sources and/or high-energy cosmological phenomena such as inflation, cosmic strings, and phase transitions (for a review, see, e.g., \cite{2000PhR...331..283M}). In particular, recent detections of compact binary coalescences \cite{2016PhRvL.116f1102A, 2016PhRvL.116x1103A,2016PhRvX...6d1015A, 2017PhRvL.118v1101A, 2017ApJ...851L..35A, 2017PhRvL.119n1101A} suggest that the expected number of such events over the cosmological scales would be too large to be individually resolved, and low signal-to-noise ratio events can contribute to the stochastic GW, which would be potentially detectable with currently operating ground-based detectors \cite{2018PhRvL.120i1101A} (see also \cite{2011RAA....11..369R, 2011ApJ...739...86Z, 2011PhRvD..84h4004R, 2011PhRvD..84l4037M, 2013MNRAS.431..882Z, 2015A&A...574A..58K}). 

There are, however, several concerns and issues toward a decisive detection of stochastic GWs. One of the most serious concerns is the contamination by the correlated environmental noise between detectors. Because of the random nature of the signal, a detection of stochastic GW is made possible by taking a cross-correlation between the data streams obtained from the multiple sets of detector. This cross-correlation technique offers a way to isolate the stochastic GW signals from detector's noise if the noises are totally uncorrelated. It has been pointed out, however, 
by Refs.~\cite{1992PhRvD..46.5250C, 1999PhRvD..59j2001A} that the correlated noise arises from the (stationary) global electromagnetic fields on the Earth, known as Schumann resonances \cite{1952ZNatA...7..149S, 1952ZNatA...7..250S}, through the coupling with magnets or magnetically susceptible materials in the laser interferometer system. Such a correlated noise can also give an impact on searches for transient GW events \cite{2017CQGra..34g4002K}. 

There are thus several experimental and theoretical studies to estimate the impact of correlated noise, and a technique to mitigate its impact has been also proposed in Refs.~\cite{2013PhRvD..87l3009T, 2014PhRvD..90b3013T, 2017PhRvD..96b2004H}. While the recent study suggests that the correlated noise budget including the one arising from the Schumann resonance is less significant and ignorable at LIGO detectors \cite{2017PhRvL.118l1101A}, the potential impact still remains at other detector sites, and a clear signal of Schumann resonances has been indeed detected through the magnetometer measurements between Virgo \cite{2015CQGra..32b4001A} and KAGRA \cite{2012CQGra..29l4007S} sites \cite{2018PhRvD..97j2007C}. 

In this respect, we have recently presented a simple analytical model to estimate the impact of correlated magnetic noise \cite{2017PhRvD..96b2004H}. The model reproduces the major trend of the measured global correlation between the GW detectors via magnetometer, and the estimated value of the impact of correlated noise quantitatively matches those inferred from the measurement results. Then, as an implication, we have explored the possible impact of the correlated noise on the detection of stochastic GWs from existing four detectors and planned detector, LIGO India \cite{2013IJMPD..2241010U}, finding that in the pessimistic case that most of the detector pairs are completely dominated by the correlated noise, LIGO Hanford-Virgo and Virgo-KAGRA pairs would be possibly less sensitive to the correlated noise, and may achieve the best sensitivity to the stochastic GWs.

While the analytic model in our previous paper shows several interesting properties, and can even be used to quantitatively estimate the impact of correlated noises, several simplifications made in the model need to be verified and/or scrutinized for a proper modeling of the correlated magnetic noise. Apart from the nonstationarity of the Schumann resonances, one potentially important effect may be the inhomogeneous distribution of the exciting sources of global magnetic fields, which can result in the anisotropies of the magnetic field spectrum. Although our previous study simply assumed the isotropic distribution of exciting sources, the lightning sources are in reality associated with global weather activity, and these are known to concentrate on the continental areas in the tropics. Indeed, the time variation of Schumann resonances measured at widely separated radio stations clearly suggests the inhomogeneous distribution of exciting sources for magnetic fields \cite{2009JASTP..71.1405S, 2011SGeo...32..705S}. The anisotropic magnetic field induced by the inhomogeneous lightning distribution may alter the correlated properties of magnetic noises, and can add another impact on the detection of stochastic GWs.

The primary purpose of this paper is to present a comprehensive study of the impact of correlated magnetic noise on the detection of stochastic GWs, taking the anisotropic distribution of lightning sources into consideration. For this purpose, we extend our analytical model to include the anisotropies in the magnetic field spectrum.  Meanwhile, the impact of the correlated magnetic noise crucially depends on the underlying assumption of GWs. While we previously focused on the unpolarized tensor GWs, there may be a possibility to have nonstandard polarization modes, and the detection of such GWs would be important. We will thus discuss quantitatively how the correlated magnetic noise can give an impact on the detectability of nonstandard GWs.

This paper is organized as follows.  In Sec.~\ref{sec:formulation}, after briefly reviewing the cross-correlation analysis, we introduce the analytical model that has been proposed in our previous paper. In Sec.~\ref{sec:special_case}, taking the anisotropies in the lightning source distribution seriously, we consider an extension of our analytical model, and discuss its potential impact on the detection of stochastic GWs. Section.~\ref{sec:realistic} then presents a quantitative estimation of the impact of correlated magnetic noise. Based on the observed spatial distribution of lightning activities, we evaluate the size of correlated magnetic noise for each pair of ongoing and upcoming second-generation GW detectors, and examine how the presence of anisotropies in the lightning source distribution, or equivalently the magnetic field spectrum can change the results, depending on which types of GW we observe. In Sec.~\ref{sec:discussion}, to understand the behaviors seen in the previous section, we consider a somewhat artificial setup, and discuss the geographical dependence of the impact of correlated magnetic noise. Finally, Sec.\ref{sec:conclusion} presents a summary of our important findings and conclusion. 
%
%
\section{cross-correlation analysis in the presence of correlated magnetic noise}
\label{sec:formulation}

In this section, we begin by briefly reviewing the standard cross-correlation analysis. We then consider the correlated magnetic noise, and introduce an analytical model presented in Ref.~\cite{2017PhRvD..96b2004H} that describes the coherence properties of the magnetic noise. 

\subsection{Cross-correlation analysis}
\label{sec:cross_correlation}

Let us first denote the time-series output data at the $i$th detector by
\begin{align}
 s_{i}(t)=h_{i}(t)+n_{i}(t),
\label{eq:output_data}
\end{align}
where $h_{i}$ is the strain amplitude produced by stochastic GWs and $n_{i}$ is the noise strain.
Since the signal of stochastic GWs is supposed to be very weak and have random properties, it is hard to distinguish between the GW signals and instrumental noise only with a single detector. One way to discriminate the GW signal from the noise is to use multiple sets of detectors and to cross-correlate between the output data. 
 
Given the two output data with the observation time of $T$, we define the cross-correlation statistic $S$ as 
\begin{align}
 S=\int_{-T/2}^{T/2}dt \int_{-T/2}^{T/2}dt'\, s_{1}(t)s_{2}(t') Q(t-t').
\label{eq:correlation_signal}
\end{align}
Here, the filter function $Q$ is introduced to enhance the detectability of the GW signals, and its explicit form will be given later as the Fourier transform $\widetilde{Q}$ [see Eq.~(\ref{eq:filter_function})].

Consider first the case that the noises between two detectors are statistically uncorrelated. The expectation value of the statistic $S$  then leads to 
\begin{align}
 \langle S \rangle  =\langle S_{\rm G}\rangle,
\label{eq:cross_correlation}
\end{align}
with $\langle S_{\rm G}\rangle$ given by
\begin{align}
\langle S_{\rm G} \rangle \equiv \int_{-T/2}^{T/2}dt \int_{-T/2}^{T/2}dt'\, 
\langle h_{1}(t)h_{2}(t') \rangle \,Q(t-t') \,.
\label{Sgw}
\end{align}
Assuming that the support of the filter function in the time domain is small enough compared to the observation time, it is expressed in the Fourier domain as (e.g., \cite{1999PhRvD..59j2001A, 2009PhRvD..79h2002N})
\begin{align}
 \langle S_{\rm G} \rangle = \frac{3H_{0}^{2}}{10\pi^{2}}T
\int_{0}^{\infty}\,df\, f^{-3}\, \, \sum_{A}\Omega_{\rm gw}^{A}(f)\, \gamma_{1 2}^{A}(f) 
\,\widetilde{Q}(f)\,,
\label{eq:S_G_Fourier}
\end{align}
where $H_0=100\,h\,$km\,s$^{-1}$\,Mpc$^{-1}$ is the Hubble parameter and the function $\widetilde{Q}$ is the Fourier transform of the optimal filter function.  Here, the summation with respect to $A$ runs over the polarization modes of GWs. The quantity $\Omega_{\rm gw}^{A}$ is the normalized logarithmic energy density of the stochastic GWs, and $\gamma_{1 2}^{A}$  is the overlap reduction function which represents the coherence of the gravitational strains between the two separated detectors.

In general relativity, the tensor mode is only allowed for the polarization mode of GWs, but it would be possible in the general metric theory of gravity to have additional polarization modes, i.e., vector and scalar modes (e.g., \cite{1973PhRvL..30..884E, 1973PhRvD...8.3308E, 2014LRR....17....4W}). A measurement of vector and/or scalar polarizations thus offers an important test of general relativity. In this paper, we consider the vector and scalar modes, assuming that these are unpolarized. For the tensor modes, we examine both the unpolarized and circularly polarized GWs, as the possibility to generate polarized GWs has been pointed out by several works \cite{2006PhRvL..96h1301A, 2008PhRvD..77b3526S}. Detectability of the stochastic GWs for unpolarized vector/scalar modes and the circularly polarized tensor mode has been previously studied in both ground- and space-based detectors (see e.g., \cite{2009PhRvD..79h2002N, 2010PhRvD..81j4043N} for unpolarized vector and scalar modes, \cite{2007PhRvD..75f1302S, 2007PhRvL..99l1101S, 2008PhRvD..77j3001S} for circularly polarized tensor mode).

In the weak-signal limit (i.e. $|h_i|\ll| n_i|$), in contrast to $S$, dispersion of the cross-correlation statistic $S$, defined by $\sigma^{2} \equiv \langle S^{2} \rangle-\langle S \rangle^{2}$, is dominated by the detector's noise. Thus, one can define the signal-to-noise ratio of the  measured stochastic GWs as 
\begin{align}
 {\rm SNR_{G}} \equiv \frac{\langle S_{\rm G} \rangle}{\sigma}.
\label{eq:SNR_G}
\end{align}
Under the assumption that noises follow the Gaussian statistics, the quantity $\sigma$ is expressed as
\begin{align}
\displaystyle{\sigma^{2} \simeq \frac{T}{2}
\int_{0}^{\infty}\,df\,P_{1}(f)\,P_{2}(f)\,|\widetilde{Q}(f)|^{2}},
\label{eq:variance}
\end{align}
where the function $P_i$ is the instrumental noise spectrum for $i$th detector. Maximizing the  signal-to-noise ratio then leads to the optimal filter in the following form \cite{1999PhRvD..59j2001A}: 
\begin{align}
 \displaystyle{\widetilde{Q}(f)\propto\frac{\sum_{A}\Omega_{\rm gw}^{A}(f)\, \gamma_{1 2}^{A}(f) }{f^{3}\,P_{1}(f)\,P_{2}(f)}}\,.
\label{eq:filter_function}
\end{align}

Provided the template of the stochastic GW spectrum $\Omega_{\rm gw}^A$, Eq.~(\ref{eq:SNR_G}) quantifies the  statistical significance of measured GW signals. However, 
the crucial assumption in the standard cross-correlation analysis is that the noises are statistically uncorrelated. In what follows, we consider the situation where the mirror control system in the laser interferometers is coupled to the global magnetic fields in the Earth-ionosphere cavity to some extent, and this leads to a certain amount of statistically correlated noises.   

Let us decompose the strain amplitude of the noise $n_{i}$ in Eq.~(\ref{eq:output_data}) into two pieces: 
\begin{align}
 n_{i}(t)=n_{i}^{\rm I}(t)+n_{i}^{\rm B}(t).
\end{align}
Here, $n_{i}^{\rm I}(t)$ is the instrumental noise originated from local disturbances, and $n_{i}^{\rm B}(t)$ represents the correlated noise induced by the global magnetic fields on the Earth. In the presence of the second term, the expectation value of the cross-correlation statistic becomes $\langle S\rangle=\langle S_{\rm G}\rangle+\langle S_{\rm B}\rangle$ with $\langle S_{\rm B}\rangle$ given by
\begin{align}
 \langle S _{\rm B} \rangle = 
\int_{-T/2}^{T/2}dt\int_{-T/2}^{T/2}dt'\, \langle n_{1}^{\rm B}(t)n_{2}^{\rm B}(t') \rangle \,Q(t-t'). 
\label{eq:sb}
\end{align}

Since the coupling between the mirror control system and the global magnetic field is supposed to be small, we may consider that the correlated noise $n_i^{\rm B}$ is linearly proportional to the global magnetic field $B^a$ at the $i$th detector's position, ${\bm x}_i$. Then, the correlated noise is generally expressed  in terms of quantities in the Fourier domain as follows:
\begin{align}
\widetilde{n}_{i}^{\rm B}(f)=r_{i}(f)\,\left[{\widehat{\bm X}}_{i}\,\cdot \widetilde{{\bm B}}(f,{\bm x}_{i})\right].
\label{eq:conv_noise}
\end{align}
Here, the frequency-dependent quantity $r_i$ is called the transfer function, which characterizes the strength of the coupling between the detector and 
magnetic field, and the unit vector $\widehat{\bm X}_i$ describes the directional dependence of its coupling.
With Eq.~(\ref{eq:conv_noise}),
the expectation value $\langle S _{\rm B} \rangle$ is rewritten in terms of the quantities in the Fourier domain with
\begin{align}
\langle S _{\rm B} \rangle 
&= T\, \int_{0}^{\infty}\,df \,\,{\rm Re}\bigl[r_{1}^{\ast}(f)\,r_{2}(f)\,M_{1 2}(f)\bigr]\,\widetilde{Q}(f),  
\label{eq:snrBB2}
\end{align}
where the function $M_{12}$ is the correlated magnetic noise spectrum for a pair of detectors, given by 
\begin{align}
M_{12}(f)&= \widehat{X}_{1,a}\widehat{X}_{2,b} \,\,
\langle \widetilde{B}^{a *}(f,{\bm x}_{1}) \widetilde{B}^b(f',{\bm x}_{2})\rangle',
\label{eq:M_12}
\end{align}
where the prime in $\langle \cdots \rangle '$ denotes that we removed the delta function $\delta_{\rm D}(f-f')$. The labels $a,\,b$ run from $1$ to $3$. 

$M_{12}$ is one of the important quantities that determines the magnitude of the correlated noise, and as pointed out in Ref.~\cite{2017PhRvD..96b2004H}, this quantity depends not only on the strength and propagation of magnetic fields, but also on the coherence of the response of the two separated detectors to the magnetic fields.  In next subsection, we present a simple analytical model of $M_{12}$ from Ref.~\cite{2017PhRvD..96b2004H}.

\subsection{Correlated magnetic noise spectrum}
\label{sec:magnetic_model}
To investigate further the impact of correlated magnetic noise, we follow Ref.~\cite{2017PhRvD..96b2004H}, and consider a simple analytical model. Note that even with a simplified setup, the model captures several important properties of correlated magnetic noise that have been partly observed through the measurement with magnetometers \cite{2013PhRvD..87l3009T,2014PhRvD..90b3013T}. The basic assumptions of the model are  summarized as follows:
\begin{enumerate}
\item The magnetic noise spectrum describes the frequency-dependent coherence of the global magnetic field between two detectors, which is expressed as a sum of the discrete Schumann resonance modes convolving the line-shape function. Here, the Schumann resonances are idealistically represented by a superposition of the axisymmetric transverse magnetic (TM) modes of the Earth-ionosphere cavity with respect to each exciting source \cite{1998clel.book.....J}.  
\item The TM modes are generated by lightning sources which are produced continuously in a stationary random process. That is, the amplitude of the TM mode, ${\widetilde B}$, has a random nature, and is characterized by the power spectrum, which is the function of the angular position of lightning source $\widehat{\bm \Omega}$ and frequency $f$. For simplicity, we further assume that the lightning distribution is isotropic.  
\end{enumerate}

Based on the above assumptions, 
the explicit expression for the correlated magnetic noise spectrum $M_{12}$ is given by Ref.~\cite{2017PhRvD..96b2004H}
\begin{align}
 M_{1 2}(f) =\frac{1}{8\pi}{P_{\rm B}(f)}\,\,\sum_{\ell} \frac{|E_\ell(f)|^{2}}{|E_\ell(f'_{\ell})|^{2}}\,
\gamma_{\ell}^{\rm B}(\widehat{\bm r}_{1},\widehat{\bm r}_{2}).
\label{eq:mij}
\end{align}
Here, $P_{\rm B}$ is the (single-sided) power spectrum density of the magnetic field, and is defined through the ensemble average of the amplitude of the TM mode $\widetilde{B}$, given by (assuming isotropic distribution)
\begin{align}
 \langle {\widetilde B}^{*}(f,\widehat{\bm \Omega})\,{\widetilde B}(f', \widehat{\bm \Omega}')\rangle
= \frac{\delta_{\rm D}^{2}(\widehat{\bm \Omega},\widehat{\bm \Omega}')}{4\pi} \,\delta_{\rm D}(f-f')\frac{P_{\rm B}(f) }{2}.
\label{eq:expectation_valueB}
\end{align}
Here, the shape function $|E_\ell(f)|^{2}|$ is given by \cite{1998clel.book.....J}
\begin{align}
|E_\ell(f)|^{2} \propto \frac{1}{(f-f'_{\ell})^{2}+\{f_{\ell}/(2{\cal Q})\}^{2}}, 
\label{eq:spectral_form}
\end{align}
where the eigenfrequency $f'_\ell$ is the observed resonance frequency, which is slightly shifted to the one in the idealistic case, $f_\ell$, by a factor of $0.78$, i.e., $f_\ell'=0.78f_\ell$, arising from several reasons including the imperfect conductivity of the Earth-ionosphere cavity.  The quantity ${\cal Q}$ is the so-called quality factor, for which we set ${\cal Q}=5$, close to the one inferred from the observed spectrum of Schumann resonances (e.g., \cite{2013PhRvD..87l3009T,1998clel.book.....J,1983JATP...45...55S,2007quality}).  

In Eq.~(\ref{eq:mij}), $\gamma_{\ell}^{\rm B}$, which characterizes the coherence of the global magnetic field, is analytically expressed in the case of axisymmetric TM modes as follows:
\begin{align}
 \gamma_{\ell}^{\rm B}(\widehat{\bm r}_{1},\widehat{\bm r}_{2}) &= 
\frac{(2\ell+1)}{2\pi}\frac{(\ell-1)!}{(\ell+1)!} \nonumber \\
\times &
\int_{S^{2}}\,d^2{\widehat {\bm \Omega}} 
\,{\cal P}_{\ell}^{1}(\widehat{\bm \Omega} \cdot \widehat{\bm r}_{1})
\,{\cal P}_{\ell}^{1}(\widehat{\bm \Omega} \cdot \widehat{\bm r}_{2})\nonumber \\
\times & 
\, \{{\widehat{\bm e}}_{1}(\widehat{\bm \Omega}) \cdot \widehat{\bm X}_{1}\}\,
\, \{{\widehat{\bm e}}_{2}(\widehat{\bm \Omega}) \cdot \widehat{\bm X}_{2}\}\,,
\label{eq:gamma_integral}
\end{align} 
where the function ${\cal{P}}_{\ell}^1$ is the associated Legendre polynomials.
$\widehat{\bm r}_{i}$ is the unit vector pointing from the Earth's center to the $i$th detector position, $\widehat{\bm e}_i$ points to the azimuthal direction with respect to each lightning source and is given by
\begin{align}
&\widehat{\bm e}_i(\widehat{\bm \Omega})=\frac{\widehat{\bm \Omega}\times\widehat{\bm r}_i}{|\widehat{\bm \Omega}\times\widehat{\bm r}_i|}.
\label{eq:unit_vect_ei}
\end{align}

The analytical model given above accounts for the magnetic noise spectrum $M_{12}$ found in Ref.~\cite{2013PhRvD..87l3009T} through the measurement of magnetosensor, and is used in our previous paper to estimate the impact of correlated magnetic noise on the detection of stochastic GWs. We then found that 
for a given strength of the transfer function $r_i(f)$, the model can predict quantitatively the size of correlated noise, which closely matchs with those estimated by \cite{2014PhRvD..90b3013T}. Nevertheless, 
there are oversimplification and crucial assumptions in the analytical model, which have to be tested and validated toward a realistic modeling of correlated noise.
In what follows, we shall discuss one of the crucial aspects of the Schumann resonance, i.e., anisotropic distribution of lightning sources.

\begin{figure*}[tb]
\begin{center}
\includegraphics[height=7.5cm,angle=0,clip]{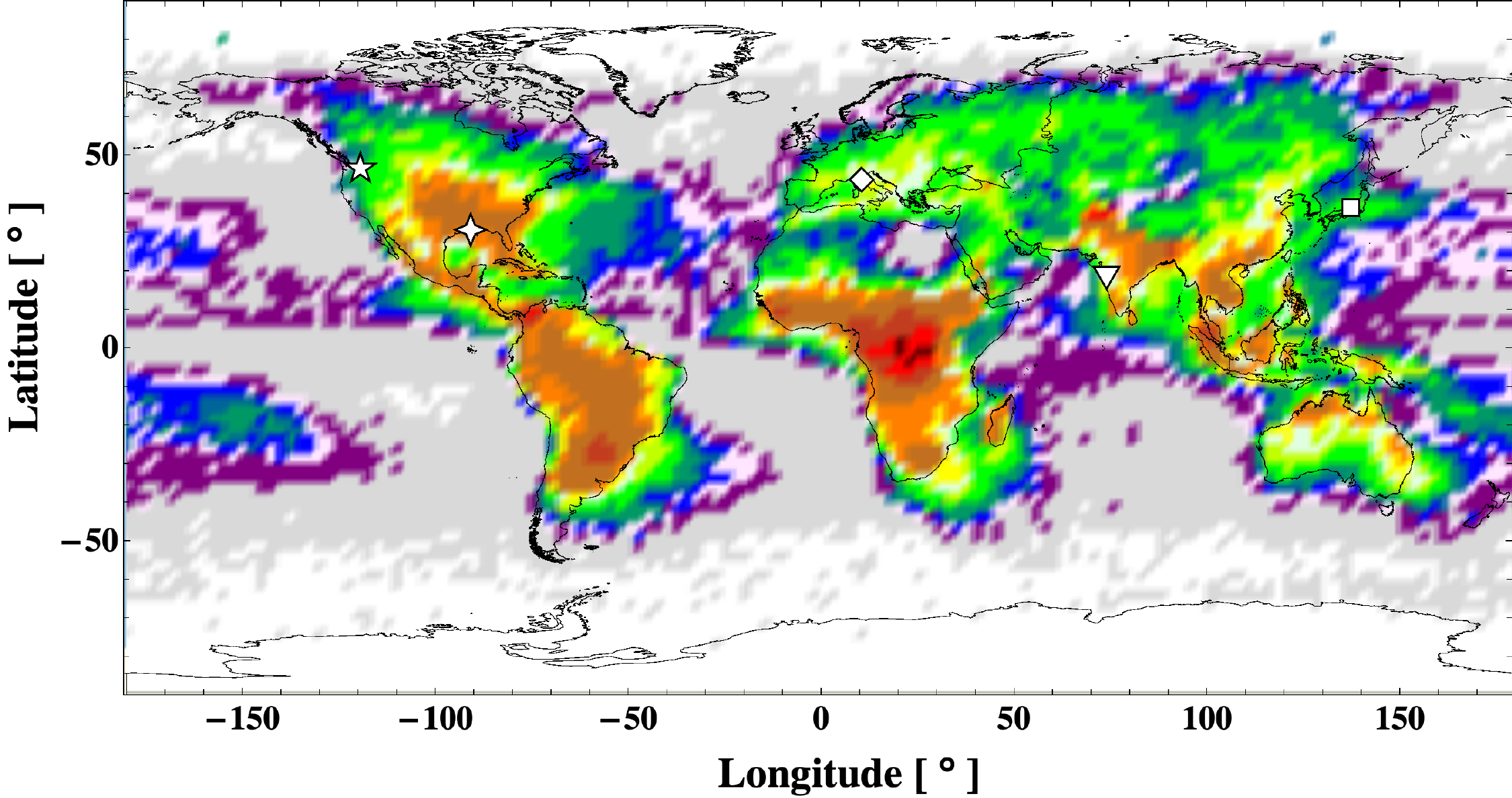}
\includegraphics[height=7.64cm,angle=0,clip]{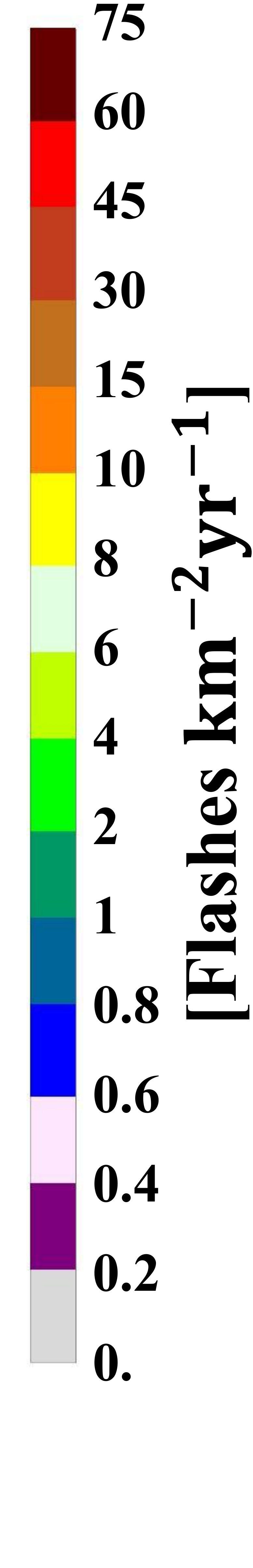}
\end{center}
\caption{World-wide density plot of the mean annual flash rate with grid size of $2.5^{\circ}\times2.5^{\circ}$. The color indicates the lightning flashes per km$^2$ per year. For reference, the location of the second-generation laser interferometers are indicated as open symbols: LIGO Hanford (five-pointed star), LIGO Livingston (four-pointed star), LIGO India (triangle), Virgo (diamond), and KAGRA (square). 
}
\label{fig:worldmap}
\end{figure*}

\section{Anisotropies in magnetic field spectrum}
\label{sec:special_case}

In this section, relaxing the assumption of the isotropic distribution of the lightning sources in the analytical model, we discuss how the presence of an anisotropic component alters the correlated noise properties.

Indeed, the distribution of lightning sources is known to be anisotropic, and rather concentrates on the continents close to the equator, associated with climate activity. 
Figure.~\ref{fig:worldmap} shows the spatial  distribution of lightning activity, which is taken from the dataset of the mean annual flash rate observed by the Optical Transient Detector and the Lightning Imaging Sensor \cite{2014AtmRe.135..404C, 2016BAMS...97.2051A}.\footnote{http://lightning.nsstc.nasa.gov/data/data$\_$lis-otd-climatology.html} The color indicates the number of flashes per km$^2$ per year. Higher lightning activity is found at the equator of the African continent. The second-generation laser interferometers, depicted as open symbols, are all located at the northern hemisphere, where the lightning activity is less significant. However, due to the spatial inhomogeneities of the lightning activity, the spectrum of magnetic fields, represented as a superposition of lightning-induced magnetic fields, can become anisotropic, and thus the actual size of the impact of correlated magnetic noise would differ from each other, largely depending on their geographical location. 

To see how the correlated noise properties will be changed, we extend the analytical model to incorporate the anisotropies in the magnetic field spectrum.  In what follows, the magnetic field spectrum $P_{\rm B}$ defined at Eq.~(\ref{eq:expectation_valueB}) is generalized to allow the anisotropic component, and is considered to be a function of not only  frequency $f$ but also angular position $\widehat{\bm \Omega}$, i.e., $P_{\rm B}(f,\widehat{\bm \Omega})$. Then, the magnetic noise spectrum, $M_{12}$, has to be modified, and the spectrum $P_{\rm B}$ in Eq.~(\ref{eq:mij}) is replaced with the sky-averaged spectrum, $\overline{P}_{\rm B}(f)=(4\pi)^{-1}\int d^2\widehat{\bfOmega}\,P_{\rm B}(f, \widehat{\bfOmega})$. The anisotropies in the magnetic field spectrum appear in the coherence function $\gamma^{\rm B}$, which  is now given by
\begin{align}
& \gamma_{\ell}^{\rm B}(\widehat{\bm r}_{1},\widehat{\bm r}_{2}) = 
\frac{(2\ell+1)}{2\pi}\frac{(\ell-1)!}{(\ell+1)!} \nonumber \\
\times &
\int_{S^{2}}\,d^2{\widehat{\bfOmega}} 
\, \frac{P_{\rm B}(f,\widehat{\bm \Omega})}{\overline{P}_{\rm B}(f)}
\,{\cal P}_{\ell}^{1}(\widehat{\bm \Omega} \cdot \widehat{\bm r}_{1})
\,{\cal P}_{\ell}^{1}(\widehat{\bm \Omega} \cdot \widehat{\bm r}_{2})\nonumber \\
\times & 
\, \{{\widehat{\bm e}}_{1}(\widehat{\bfOmega}) \cdot \widehat{\bm X}_{1}\}\,
\, \{{\widehat{\bm e}}_{2}(\widehat{\bfOmega}) \cdot \widehat{\bm X}_{2}\}\,.
\label{eq:gamma_integral2}
\end{align} 
That is, on top of the geometry of the detector's pair and coupling parameters $\widehat{\bm X}_i$, $\gamma_\ell^{\rm B}$  has an additional dependence on the anisotropies of the magnetic field spectrum.

Before going to a quantitative study in Sec.~\ref{sec:realistic}, it would be helpful to see how the presence of anisotropies qualitatively changes the properties of $\gamma^{\rm B}_\ell$. Let us recall in the case of the isotropic magnetic field spectrum that irrespective of the geometric configuration of the detector's pair, there exists a certain set of projection vectors $\widehat{\bm X}_i$ that cancel $\gamma^{\rm B}_\ell$. To be precise, the function $\gamma_\ell^{\rm B}$ vanishes if and only if the two projection vectors $\widehat{\bm X}_i$ are orthogonal each other, and one of them points to the direction parallel or perpendicular to the great circle connecting the pair of detectors. In Ref.~\cite{2017PhRvD..96b2004H}, we show that this nulling condition is solely due to the symmetric reason. Thus, in the presence of anisotropies in the magnetic field spectrum, the nulling condition is prone to be violated.

In Appendix~\ref{appendix:special}, we show that the nulling condition mentioned above still holds in the anisotropic case. However, it is only the case for a certain pair of detectors, and under the special symmetry for the magnetic field spectrum. To be precise, the magnetic field spectrum should have the axial symmetry whose symmetric axis is parallel or perpendicular to the plane spanned by the detector's position vectors pointing from the Earth's center ($\widehat{\bm r}_1$ and $\widehat{\bm r}_2$). In other words, no global nulling condition exists, and even if one can tune the coupling parameters $\widehat{\bm X}_i$, the correlated noise cannot be canceled for all pairs of detectors. This indicates that the presence of anisotropies generally worsens the situation, and the impact of correlated magnetic noise on the detection of stochastic GWs becomes more significant. 

In the next section, we will see quantitatively how the impact of correlated noise sensitively depends on the anisotropies in the magnetic spectrum.

\section{Estimation of correlated magnetic noise from anisotropic lightning sources}
\label{sec:realistic}

In this section, based on the observed lightning distribution shown in Fig.~\ref{fig:worldmap}, we quantitatively estimate the impact of correlated noise on the detection of the GW signal in the presence of anisotropies.  After summarizing our basic setup and assumptions in Sec.~\ref{sec:setup}, we present the results in Sec.~\ref{sec:results}.

\subsection{Setup}
\label{sec:setup}

As we mentioned in Secs.~\ref{sec:intro} and \ref{sec:magnetic_model}, the Schumann resonances are sourced by the lightning activity, and the spatial inhomogeneities in the lightning distribution suggests a non-negligible amount of  anisotropies in the magnetic field spectrum. That is, the angular dependence of the spectrum $P_{\rm B}$ is likely to follow the lightning distribution shown in Fig.~\ref{fig:worldmap}. Then, we introduce the parameter $\epsilon$ that characterizes the strength of anisotropies with respect to the isotropic component. Assuming that the anisotropic component is independent of frequencies, the function $P_{\rm B}(f,\widehat{\bfOmega})$ is expressed as  
\begin{align}
P_{\rm B}(f,\widehat{\bfOmega})=\overline{P}_{\rm B}(f) W(\widehat{\bfOmega})
\label{eq:P_B_anisotropies}
\end{align}
with the function $W$ given by
\begin{align}
W(\widehat{\bfOmega})=(1-\epsilon)+ \epsilon\, w(\widehat{\bfOmega}), 
\label{eq:weight-func}
\end{align}
where the anisotropic component characterized by the function $w$ is normalized as $4\pi=\int d^2\widehat{\bfOmega} \,w(\widehat{\bfOmega})$, and we assume that it  
simply follows the lightning distribution in Fig.~\ref{fig:worldmap}.\footnote{To be precise, Fig.~\ref{fig:worldmap} is given by the pixelized dataset tabulated as $(\theta_i,\phi_i,w_i)$ $(i=1,\cdots,N)$, where $\theta_i$ and $\phi_i$ are the latitude and longitude at $i$th pixel, respectively, and $w_i$ is the flash rate. Using these data, the properly normalized function $w$ is defined as follows:
\begin{align}
w(\widehat{\Omega})={4\pi} \frac{\displaystyle \sum_{i=1}^{N} w_{i} \, \delta_{\rm D}(\theta-\theta_{i})\delta_{\rm D}(\phi-\phi_{i})}{\displaystyle \sum_{j=1}^{N} w_{j}\,\sin \theta_{j}}.
\label{eq:weight-func2}
\end{align}
} For the frequency dependence of the isotropic part, $\overline{P}_{\rm B}$, we follow the discussions in Refs.~\cite{2017PhRvD..96b2004H, 1999PhRvD..59j2001A}, and adopt the power-law form: 
\begin{align}
 \overline{P}_{\rm B}(f)=A \left(\frac{f}{10 {\rm Hz}}\right)^{-0.88},
\label{eq:mg_spectrum}
\end{align}
with the normalization amplitude $A^{1/2}=5.89$\,pT\,\,Hz$^{-1/2}$.

In order to estimate the impact of correlated magnetic noise, a crucial part is the strength of the coupling between laser interferometers and global magnetic fields, characterized by the transfer function $r_i(f)$ [see Eqs.(\ref{eq:conv_noise}) or (\ref{eq:snrBB2})]. For a pair of $i$- and $j$th detectors, the impact on stochastic GWs characterized by $|\langle S_{\rm B}\rangle|$ simply scales as $r_i r_j$. Here, for illustrative purpose, we consider the same functional form as used in Refs.~\cite{2013PhRvD..87l3009T, 2014PhRvD..90b3013T,2017PhRvD..96b2004H} as a fiducial setup: 
\begin{align}
r_{i}(f)=\kappa_i  \times 10^{-23}\left(\frac{f}{10 \,{\rm Hz}}\right)^{-b_i} [{\rm strain}\,\,{\rm pT}^{-1}], \,\,\,\,(i=1,2).
\label{eq:coupling}
\end{align}
Here, adopting the length coupling used in Ref.~\cite{2014PhRvD..90b3013T}, we set the parameters $(\kappa_i, b_i)$ to $(2, 2.67)$ for all detectors. An updated calibration of the coupling function by Ref.~\cite{2019arXiv190302886T, Nguyen:2017} suggests that the amplitude of the coupling at LIGO has been substantially reduced by more than 1 order of magnitude in the latest instrumental setup. Nevertheless, the coupling to the magnetic field at other detectors, especially KAGRA and LIGO India, is still uncertain, and can be potentially large. We shall thus use the same coupling parameters as adopted previously, but, in Appendix \ref{appendix:newcoupling}, we also present the impact of correlated magnetic noise with the updated transfer function for LIGO Hanford and Livingston based on  Ref.~\cite{Nguyen:2017}.

Based on the setup above, in what follows, we consider the five second-generation detectors, i.e., LIGO Hanford (H), Livingston (L), India (I), Virgo (V), and KAGRA (K), and estimate the impact of correlated magnetic noise, assuming the flat spectra of the logarithmic energy density of stochastic GWs, $\Omega_{\rm gw}^A\propto f^0$, as our fiducial target. In Appendix \ref{appendix:astro}, we also investigate the cases with a spectral index of $2/3$ (i.e., $\Omega_{\rm gw}^A\propto f^{2/3}$), corresponding to the astrophysical GW background of binary coalescence \cite{2018PhRvL.120i1101A, 2011RAA....11..369R, 2011ApJ...739...86Z, 2011PhRvD..84h4004R, 2011PhRvD..84l4037M, 2013MNRAS.431..882Z, 2015A&A...574A..58K}. We then compute $|\langle S_{\rm B}\rangle|$ for various detector pairs. The resultant value of $|\langle S_{\rm B}\rangle|$ is translated into the amplitude of $\Omega_{\rm gw}h^2$ by equating $|\langle S_{\rm B}\rangle|$ with $S_{\rm G}$. To compute $|\langle S_{\rm B}\rangle|$ and $\langle S_{\rm G}\rangle$, the instrumental noise spectra $P_1$ and $P_2$ and overlap reduction function $\gamma_{12}^A$ involved in the optimal filter function $\widetilde{Q}$ need to be specified [see Eq.~(\ref{eq:filter_function})]. We adopt the same noise spectral density for each detector as used in our previous paper \footnote{For LIGO Hanford/Livingston/India, we use the table of numerical data published in Ref.~\cite{Shoemaker:2014}. For KAGRA and Virgo,  we use the fitting form of the noise spectra, given at Eqs.~(5) and (6) in Ref.~\cite{2012arXiv1202.4031M}, respectively.}. The expressions for the overlap reduction function for various types of polarized GWs are analytically known, and we use them to estimate quantitatively the impact for each case \footnote{For unpolarized tensor, vector and scalar modes, we use the analytical form given at Eqs.~(33)--(41) in Ref.~\cite{2009PhRvD..79h2002N}. For the circularly polarized tensor mode, we use Eq.~(8) in Ref.~\cite{2007PhRvL..99l1101S}.}, adopting the geometrical parameters summarized in Table V of Ref.~\cite{2017PhRvD..96b2004H}.

\begin{figure*}[tb]
\begin{center}
\includegraphics[width=8cm,angle=0,clip]{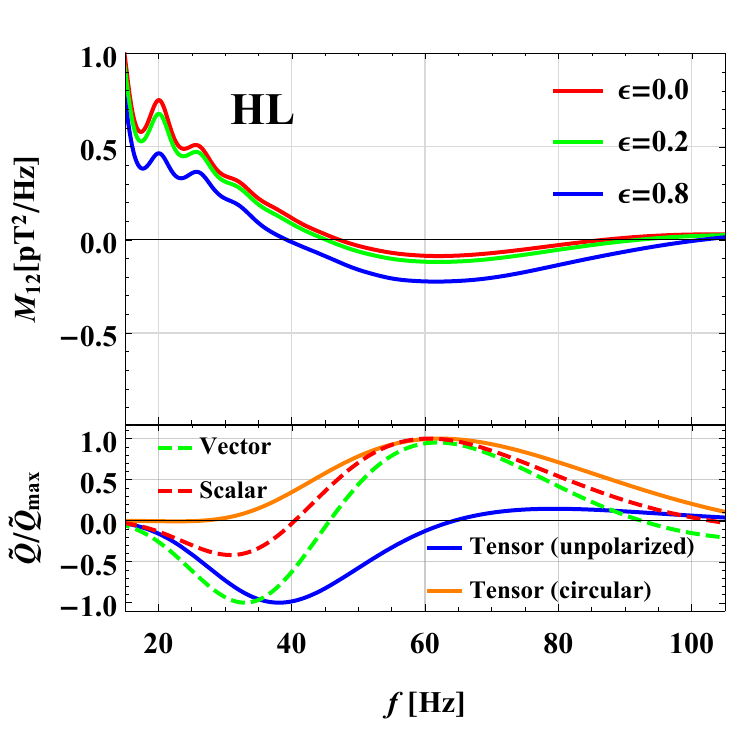}
\includegraphics[width=8.cm,angle=0,clip]{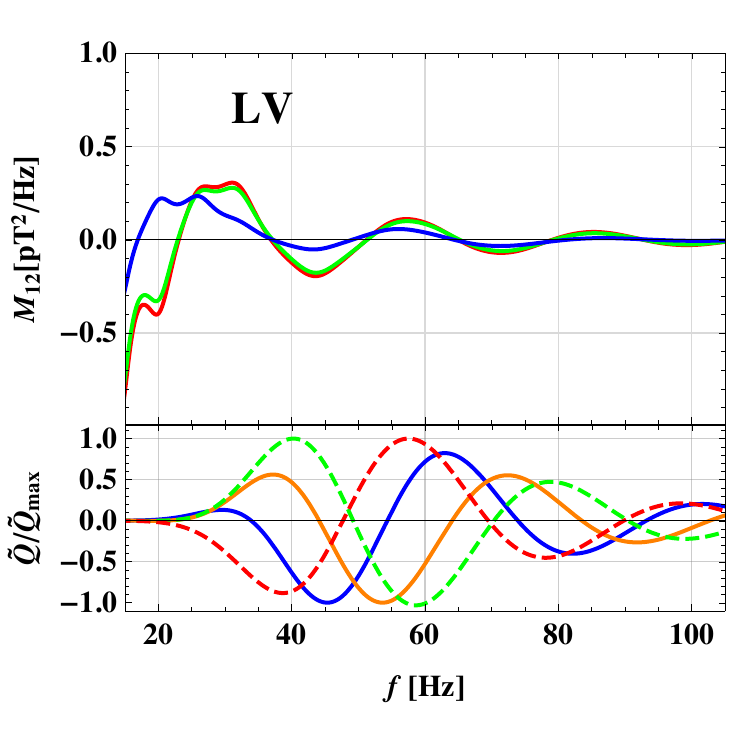}\\
\includegraphics[width=8.cm,angle=0,clip]{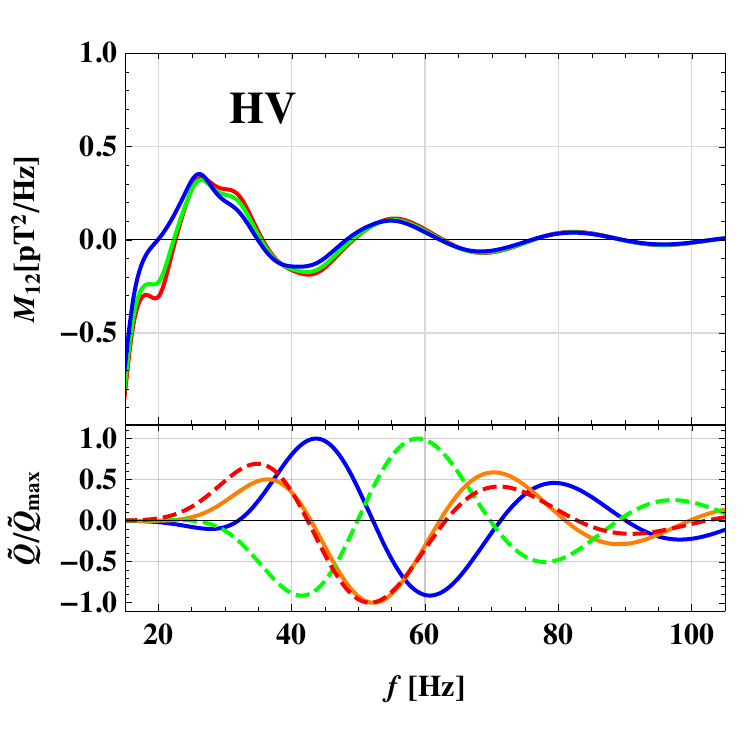}
\includegraphics[width=8.cm,angle=0,clip]{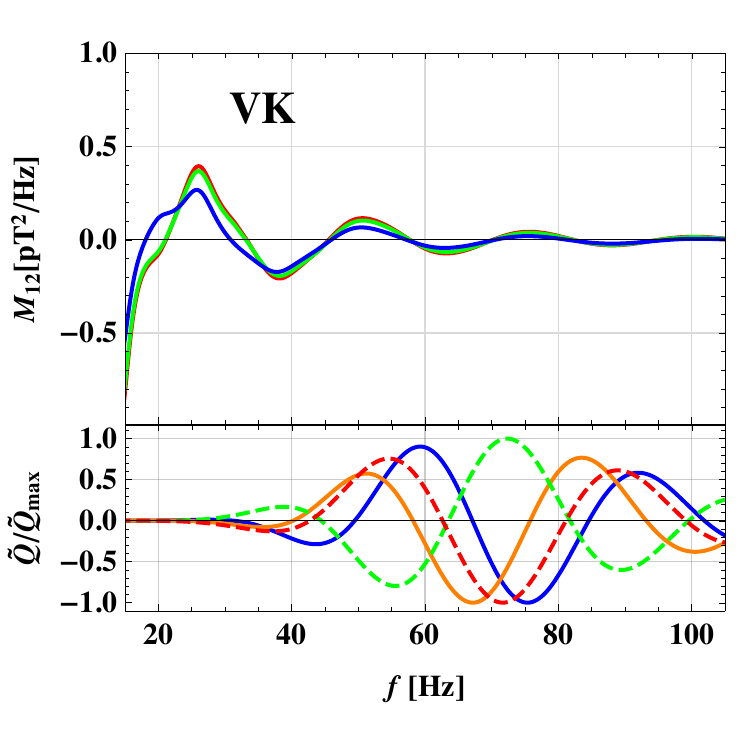}
\end{center}
\caption{Frequency dependence of the magnetic noise power spectrum, $M_{12}$ (top), and optimal filter function, $\widetilde{Q}$ (bottom), for four representative pairs: LIGO Hanford and Livingston  (HL, upper left), LIGO Livingston and Virgo (LV, upper right), LIGO Hanford and Virgo (HV, lower left), and Virgo and KAGRA (VK, lower right). The response of the magnetic noise power spectrum with respect to the parameter $\epsilon$, which characterizes the strength of anisotropies in the magnetic field spectrum, is particularly shown in different colors. Note that the projection vector, $\widehat{X}_i$, which describes the directional coupling to the GW detector, is chosen so that the cross-correlation statistic $|\langle S_{\rm B}\rangle|$ is maximized for each pair of detectors in each value of $\epsilon$. The optimal filter function is computed for various types of stochastic GWs, and  normalizing its amplitude by maximum value, $\widetilde{Q}_{\rm max}$, the results are shown in different colors and line styles. }
\label{fig:m12_compare}
\end{figure*}

\begin{figure*}[tb]
\begin{center}
\hspace{-1cm}
\includegraphics[width=8.cm,angle=0,clip]{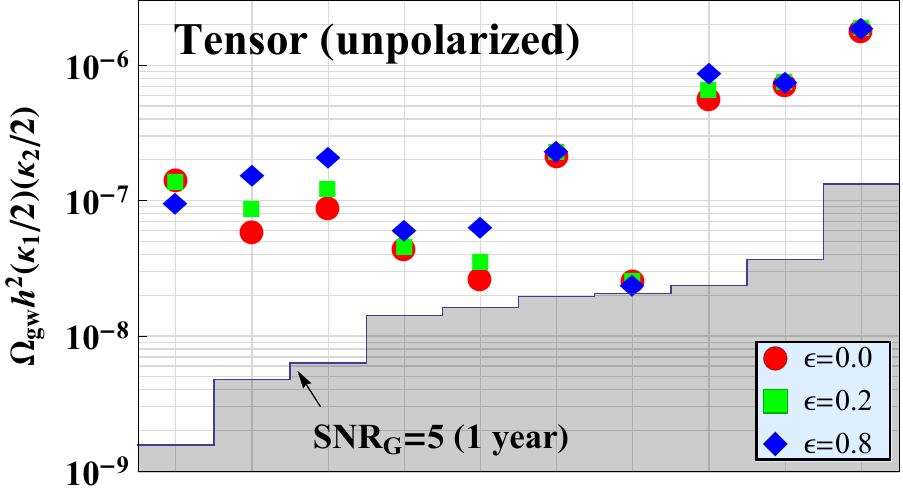}
\hspace*{5mm}
\includegraphics[width=7.4cm,angle=0,clip]{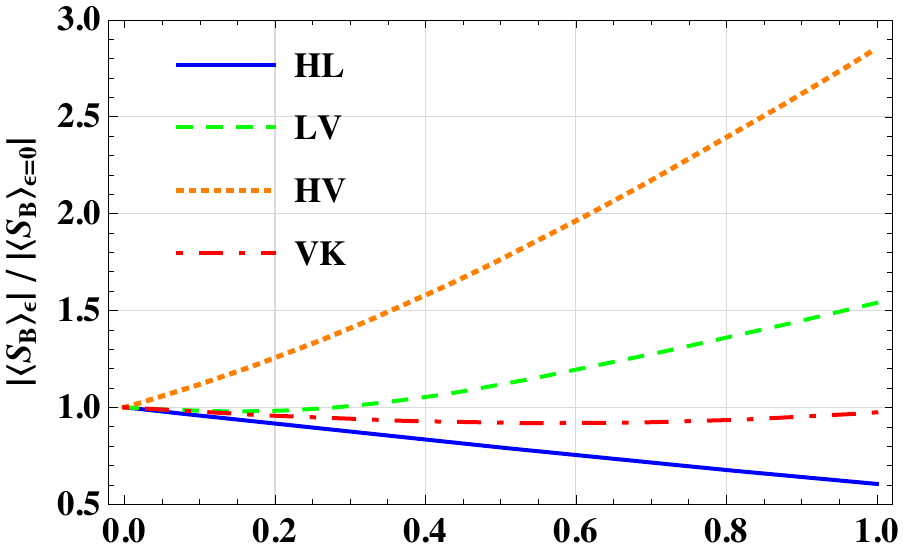}\\
\vspace*{0.0cm}
\hspace*{-29mm} 
{\bf \small HL \hspace{-0.1mm} HI \hspace{-0.1mm} LI  \hspace{-0.1mm} LV  \hspace{-0.4mm}  HV  \hspace{-1mm} VI  \hspace{-0.4mm}  VK   \hspace{-1mm} KI  \hspace{-1mm} HK  \hspace{-1mm}  LK } \hspace*{4.5cm}  $\epsilon$ \\
\vspace*{0.1cm}
\hspace*{-1cm} 
\includegraphics[width=8.cm,angle=0,clip]{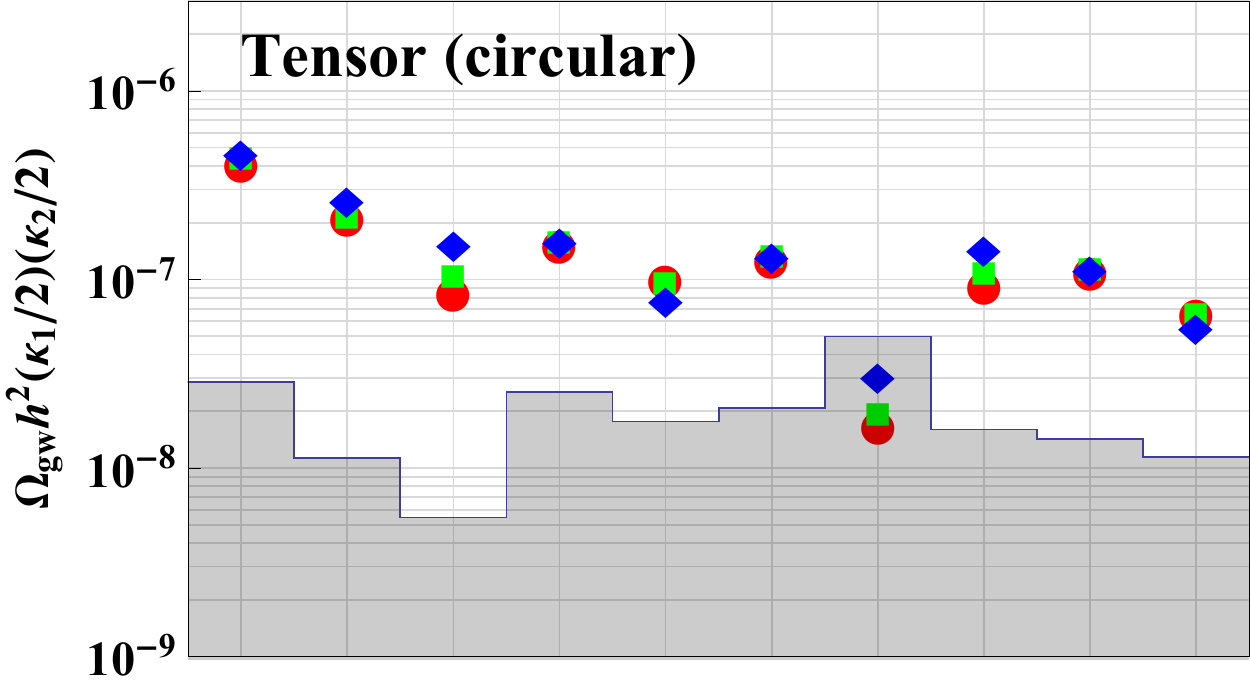}
\hspace*{5mm}
\includegraphics[width=7.4cm,angle=0,clip]{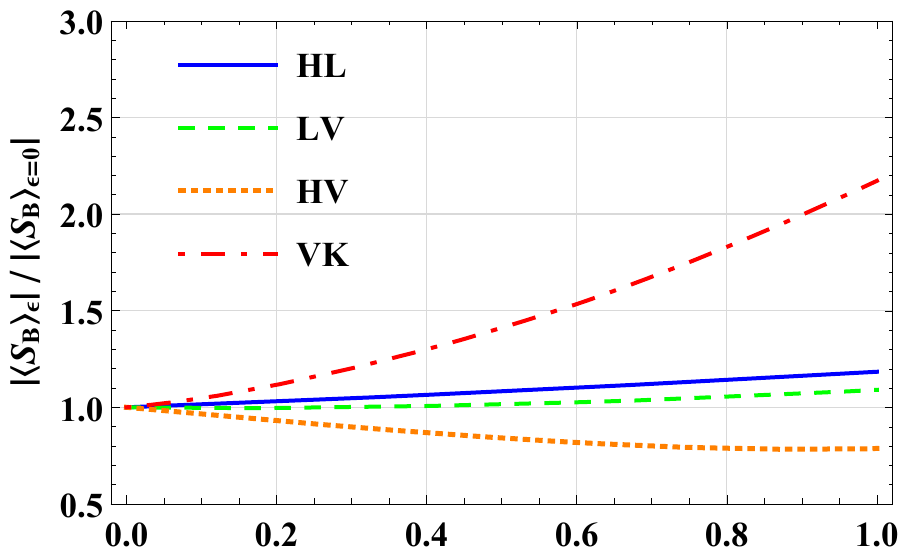}\\
\hspace*{8.5cm} $\epsilon$
\end{center}
\caption{{\it Left}: Potential impact of correlated magnetic noise on the detection of tensor GWs for $\epsilon=0$ (red circles), $0.2$ (green squares), and $0.8$ (blue diamonds). The upper panel shows the results assuming unpolarized GWs, while the lower panel plots the result for circularly polarized GWs. Here, we consider the stochastic GWs with a flat spectrum (i.e., $\Omega_{\rm gw}\propto f^0$), and the cross-correlation statistics $\langle S_{\rm B}\rangle$ are estimated for all possible combinations of detector pairs. Then, the results are translated to $\Omega_{\rm gw}h^2$ by setting $|\langle S_{\rm B}\rangle|=\langle S_{\rm G}\rangle$. Note that adjusting the projection vector, $\widehat{X}_i$, the amplitude of $|\langle S_{\rm B}\rangle|$ is maximized for each detector pair in each value of $\epsilon$. For reference, a detectable amplitude of the stochastic GWs with signal-to-noise ratio $\rm{SNR_G}=5$ is also estimated, assuming a one-year observation. Then, the region of $\rm{SNR_G}<5$ is shown in shaded color. {\it Right}: Dependence of the parameter $\epsilon$ on the correlated magnetic noise in the presence of anisotropies in the magnetic field spectrum. Normalizing their results by those in the isotropic case, we plot the ratio $|\langle S_{\rm B}\rangle_{\epsilon}|/|\langle S_{\rm B}\rangle_{\epsilon=0}|$ as function of $\epsilon$. Here, the results for four representative detector pairs are only shown for unpolarized GWs (upper) and circularly polarized GWs (lower): LIGO Hanford and Livingston (HL, blue solid), LIGO Livingston and Virgo (LV, green dashed), LIGO Hanford and Virgo (HV, orange dotted), and Virgo and KAGRA (VK, red dot-dashed). }
\label{fig:impact_flat_TITV}
\end{figure*}

\begin{figure*}[tb]
\begin{center}
\hspace{-1cm}
\includegraphics[width=8.cm,angle=0,clip]{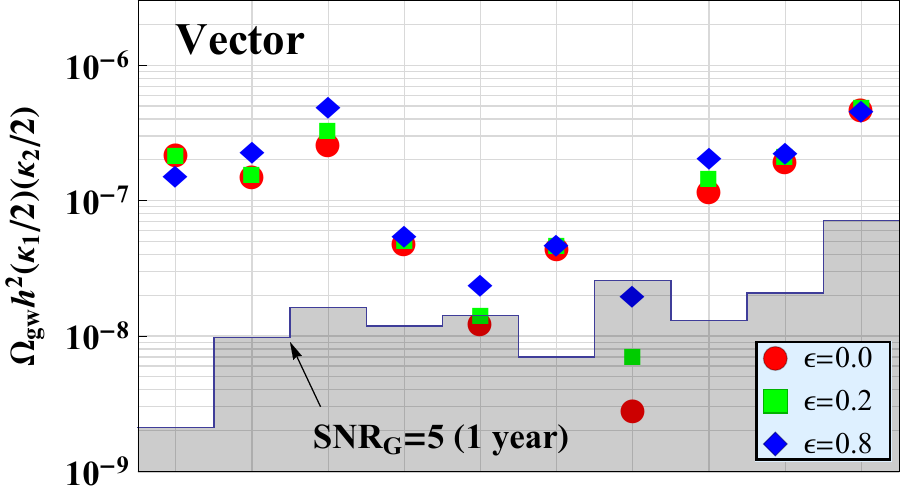}
\hspace*{5mm}
\includegraphics[width=7.4cm,angle=0,clip]{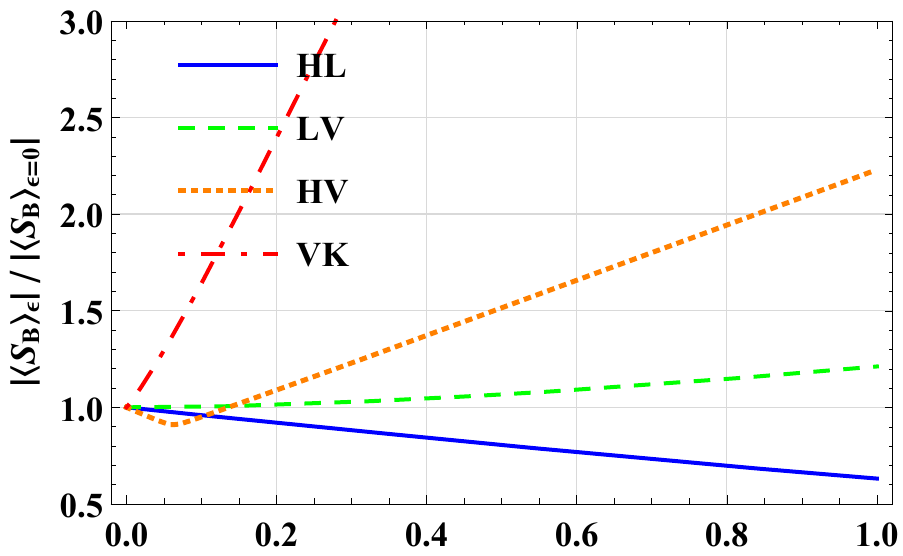}\\
\vspace*{0.0cm}
\hspace*{-29mm} 
{\bf \small HL \hspace{-0.1mm} HI \hspace{-0.1mm} LI  \hspace{-0.1mm} LV  \hspace{-0.4mm}  HV  \hspace{-1mm} VI  \hspace{-0.4mm}  VK   \hspace{-1mm} KI  \hspace{-1mm} HK  \hspace{-1mm}  LK } \hspace*{4.5cm}  $\epsilon$ \\
\vspace*{0.1cm}
\hspace*{-1cm} 
\includegraphics[width=8.cm,angle=0,clip]{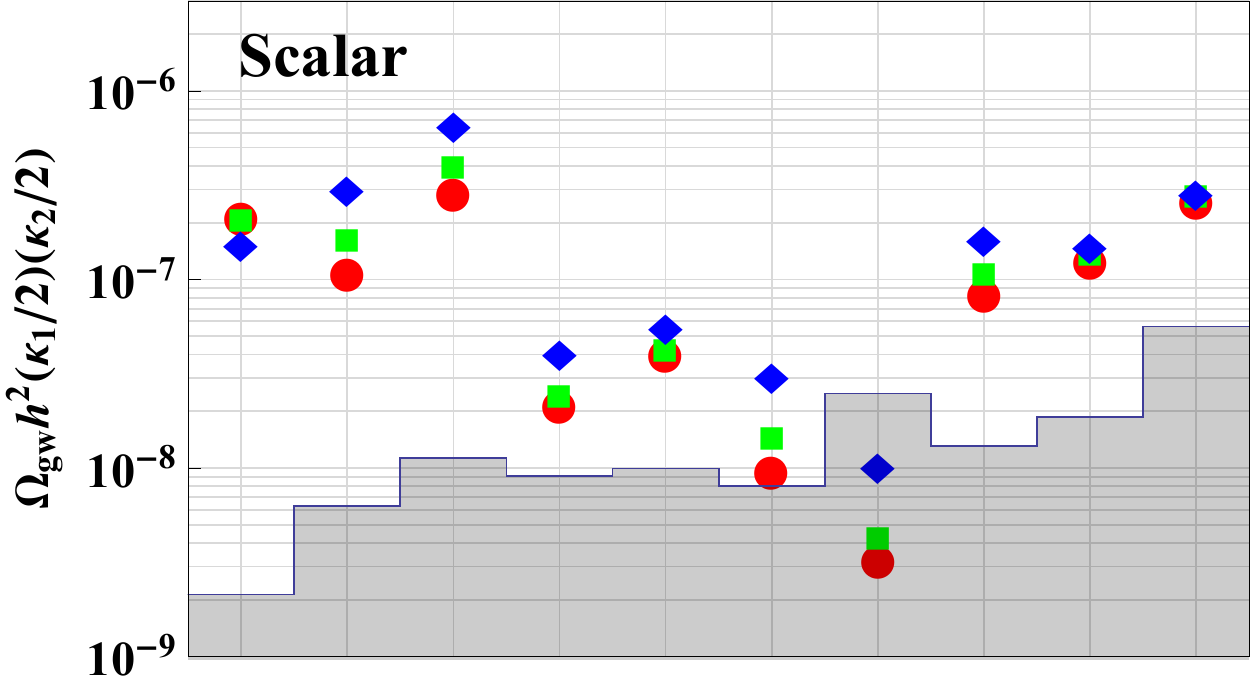}
\hspace*{5mm}
\includegraphics[width=7.4cm,angle=0,clip]{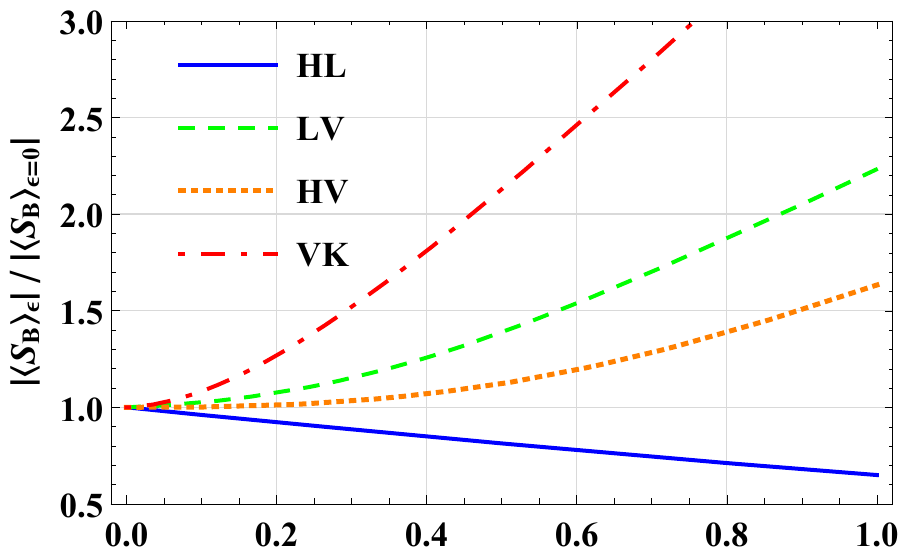}\\
\hspace*{8.5cm} $\epsilon$
\end{center}
\caption{Same as in Fig.~\ref{fig:impact_flat_TITV}, but in the case of vector (upper) and scalar (lower) GWs, assuming that the stochastic GWs are unpolarized.  
}
\label{fig:impact_flat_SV}
\end{figure*}

\subsection{Results}
\label{sec:results}

\subsubsection{Correlated magnetic noise spectrum}
\label{subsec:correlated_noise_spectrum}

Before presenting the quantitative impact, it would be instructive to first see how the presence of anisotropies in the magnetic field spectrum changes the behaviors of the correlated magnetic noise spectrum $M_{12}$, which is the key quantity to estimate the cross-correlation statistic in the presence of correlated noise, $|\langle S_{\rm B}\rangle|$.

Figure.~\ref{fig:m12_compare} shows the magnetic noise spectra $M_{12}$ for four representative pairs of interferometers, LIGO Hanford (H), Livingston (L) and Virgo (V), and KAGRA (K) detectors. Three different lines in each upper panel represent the results with different values of $\epsilon$. Here, the projection vector characterizing the direction of magnetic coupling, $\widehat{\bm X}_i$, is chosen in such a way that the cross-correlation statistic $|\langle S_{\rm B}\rangle|$ takes the maximum for each pair of detectors. 

In Fig.~\ref{fig:m12_compare}, we also plot in each panel the optimal filter function $\widetilde{Q}$ defined at Eq.~(\ref{eq:filter_function}) for various types of stochastic GWs, normalized by its maximum amplitude  $\widetilde{Q}_{\rm max}$. In all cases, as we increase the frequency, the optimal filters start to exhibit oscillatory behaviors around $f=20-100$ Hz, where the detector pair becomes sensitive to the stochastic GWs. Thus, for a nonzero $M_{12}$ around this frequency range, a large impact of the correlated magnetic noise is expected. Figure.~\ref{fig:m12_compare} indicates that the LIGO Hanford and Livingston pair seems to be sensitively affected by the correlated magnetic noise, whereas other detector pairs look less sensitive because of the rapid oscillation of both the optimal filter and magnetic noise spectrum. 

Regarding the anisotropies in the magnetic field spectrum, the changes in $M_{12}$ are basically small and are mostly coherent over $f=20-100$ Hz in all detector pairs. Hence, we naively expect its impact on the detection of stochastic GWs to vary linearly with $\epsilon$, and the variation of the impact would be a factor of 2-3. However, in the presence of oscillatory behavior in the optimal filter function, the actual size of the impact, quantified by $\langle S_{\rm B}\rangle$, is not always the case that we naively expect. As we will see below, depending on which type of GWs we observe, the impact can change by more than a factor of 2-3 for some of the detector pairs. Also, interestingly, the phase cancellation in the integrand of Eq.~(\ref{eq:snrBB2}) can happen, and the impact of correlated noise could be reduced to some extent.

\subsubsection{Impact on the detection of stochastic GWs}
\label{subsec:impact_GWs}

We now present the quantitative estimate of the impact of correlated magnetic noise varying the parameter $\epsilon$. Figures.~\ref{fig:impact_flat_TITV} and \ref{fig:impact_flat_SV} summarize the results for tensor and nontensor (i.e., vector and scalar) modes, respectively. In each figure, the left panels plot the expected amplitude of correlated magnetic noise in terms of $\Omega_{\rm gw}h^2$ for all possible pairs of detectors, which linearly scales with the coupling parameter of the transfer function, $\kappa_1\,\kappa_2$ [see Eq.~(\ref{eq:coupling})]. 
On the other hand, the right panels of Figs.~\ref{fig:impact_flat_TITV} and \ref{fig:impact_flat_SV} plot the dependence of the parameter $\epsilon$ on the cross-correlation statistic dominated by the magnetic noise, normalized by the one in the isotropic case, i.e., $|\langle S_{\rm B}\rangle_{\epsilon}|/|\langle S_{\rm B}\rangle_{\epsilon=0}|$. Here, we only show the four representative cases among all possible combinations of detector pairs: LIGO Hanford and Livingston (HL), LIGO Hanford and Virgo (HV), LIGO Livingston and Virgo (LV), and finally Virgo and KAGRA (VK).  Note that all the results shown here correspond to the most pessimistic cases in the sense that the cross-correlation statistic $\langle S_{\rm B}\rangle_{\epsilon}$ is taken to be a maximum value by adjusting the projection vectors $\widehat{\bm X}_i$ for each pair of the detectors.

Figures.~\ref{fig:impact_flat_TITV} and \ref{fig:impact_flat_SV} basically tell us that in most of the cases, the impact of correlated magnetic noise monotonically increases with the strength of anisotropies. This is what we expected. Nevertheless, a closer look at these figures reveals several nontrivial features for a specific GW mode and pairs of detectors listed below: 

\begin{itemize}
    \item For circularly polarized tensor GWs, the variation of the amplitude, $|\langle S_{\rm B}\rangle_\epsilon|$ or $\Omega_{\rm gw}h^2$, with respect to $\epsilon$ is smaller than that for other types of GWs. That is, the impact of correlated magnetic noise is mostly similar to that in the isotropic case, and it can change by a factor of 2 at maximum.    
    \item The correlated magnetic noise at the HL pair is less affected by the anisotropies in the magnetic field spectrum. This is true in all types of stochastic GWs. Interestingly, for unpolarized tensor and vector/scalar GWs, the relative impact on GWs gets suppressed the parameter $\epsilon$ increases, although the actual impact on the detection of GWs is still large. 
    \item The VK pair is the least sensitive detector pair against the correlated magnetic noise. This is indeed the case for all types of stochastic GWs, and for our fiducial setup, the VK pair achieves the best sensitivity to the GWs among all possible pairs. Note, however, that for nonstandard GWs, the VK pair is largely affected by the anisotropies in the magnetic field spectrum, and the size of impact characterized by $|\langle S_{\rm B}\rangle_\epsilon|$ or $\Omega_{\rm gw}h^2$ can change by more than a factor of $3$.  
\end{itemize}

These noticeable features basically come from the properties of the magnetic noise spectrum and optimal filter function as we have seen in Fig.~\ref{fig:m12_compare}, and are thus ascribed to the geometrical configuration of a detector pair on top of the  geographical character of the lightning activity. In the next section, we shall discuss this point in more detail, focusing on unpolarized tensor GWs.

\begin{figure*}[tb]
\begin{center}
\includegraphics[width=8.cm,angle=0,clip]{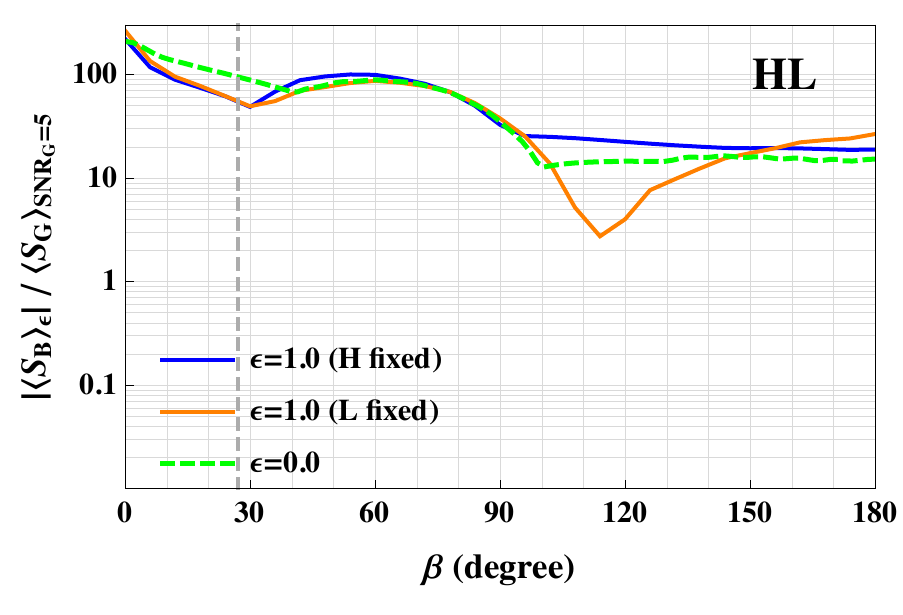}
\includegraphics[width=8.cm,angle=0,clip]{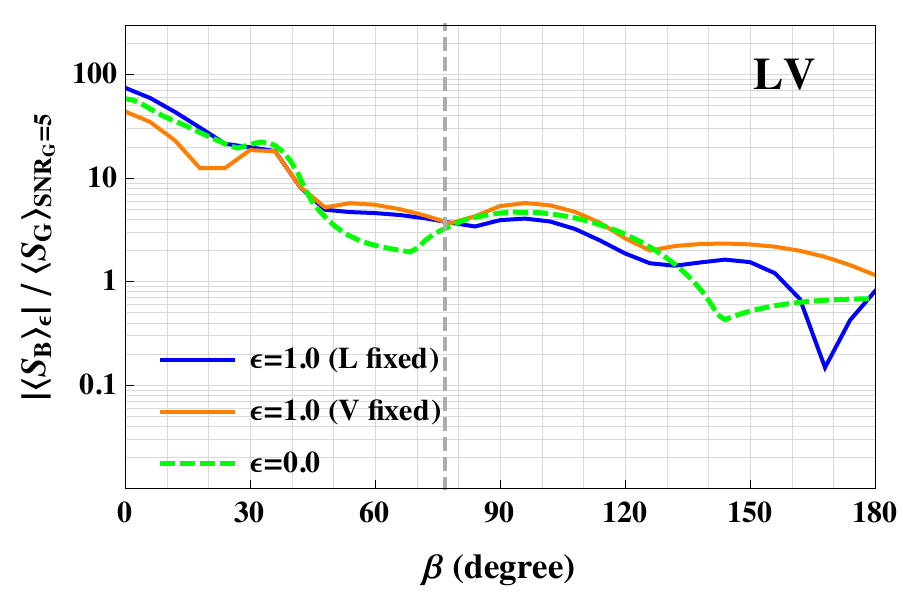}\\
\includegraphics[width=8.cm,angle=0,clip]{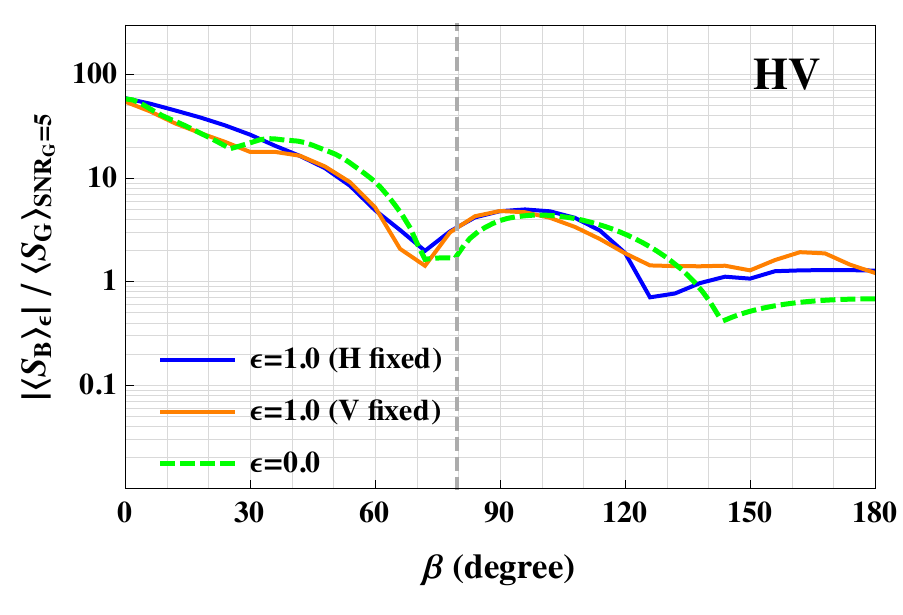}
\includegraphics[width=8.cm,angle=0,clip]{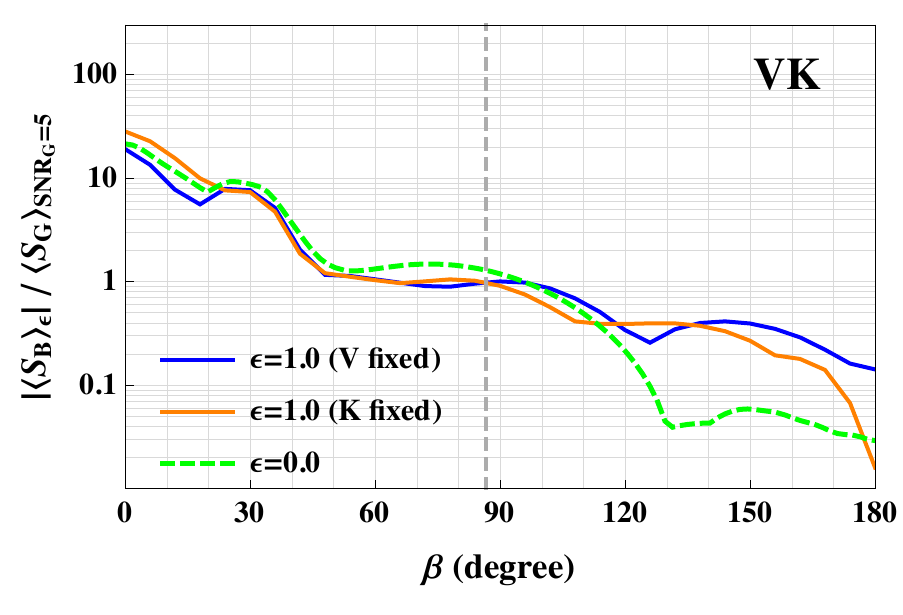}
\end{center}
\caption{Dependence of the correlated magnetic noise on the geographical setup for representative pairs of detectors: LIGO Hanford and Livingston (HL, upper left), LIGO Livingston and Virgo (LV, upper right), LIGO Hanford and Virgo (HV, lower left), and Virgo and KAGRA (VK, lower right). In each panel, shifting the location of the detector's pair along the great circle connecting their positions, we first compute the cross-correlation statistic $|\langle S_{\rm B}\rangle|$. The results are then normalized by the stochastic GW signal, $\langle S_{\rm G}\rangle$, computed with the same setup, and are plotted as a function of the separation angle, $\beta$, defined by $\beta\equiv \cos^{-1}(\widehat{\bm r}_1\cdot\widehat{\bm r}_2)$, where the unit vector $\widehat{\bm r}_i$ points to the $i$th detector position from the Earth's center. Dashed lines represent the results for $\epsilon=0$ (i.e., isotropic case), whereas solid lines are the cases with $\epsilon=1$, fixing one of the detector' s positions as indicated by the figure legend.  Note that in computing $|\langle S_{\rm B}\rangle|$, the projection vectors $\widehat{\bm X}_i$ are chosen in each pair so as to maximize $|\langle S_{\rm B}\rangle|$ in the $\epsilon=0$ case. For reference, the vertical dashed lines indicate the separation angle of the original setup. }
\label{fig:discussion}
\end{figure*}

\section{Discussion}
\label{sec:discussion}

In this section, to better understand the results obtained in Sec.~\ref{sec:results}, we consider the unpolarized GWs only, and examine how the geographical location of the detector pair changes the impact of correlated magnetic noise in the presence of the anisotropic magnetic field spectrum. For this purpose, we artificially shift the location of the detector's pair along the great circle connecting their positions, and evaluate the response of the cross-correlation statistic  $\langle S_{\rm B}\rangle_\epsilon$ to the variation of the detector's separation.  To be precise, we fix the position for one of the detector pairs to the original location, and one at another position is shifted along the great circle.\footnote{In this treatment, the orientation of the detector pair also remains unchanged with respect to the great circle.} Then, we compute $\langle S_{\rm B}\rangle_\epsilon$, and the results are plotted as a function of the separation angle, $\beta$, defined by $\beta\equiv\cos^{-1}(\widehat{\bm r}_1\cdot\widehat{\bm r}_2)$, with the unit vector $\widehat{\bm r}_i$  pointing to the $i$th detector position from the Earth's center. Note that in computing $\langle S_{\rm B}\rangle_\epsilon$, the directional coupling vector $\widehat{\bm X}_i$ is chosen in such a way that $\langle S_{\rm B}\rangle_\epsilon$ takes the maximum value in the $\epsilon=0$ case. 

Figure.~\ref{fig:discussion} shows the results for the four representative pairs: LIGO Hanford and Livingston (HL), LIGO Livingston and Virgo (LV), LIGO Hanford and Virgo (HV), and Virgo and KAGRA (VK). Here, the correlated magnetic noise, $|\langle S_{\rm B}\rangle_\epsilon|$, is normalized in each case by the cross-correlation statistic of the stochastic GW signals, $\langle S_{\rm G}\rangle$, whose amplitude is determined by the signal-to-noise ratio of ${\rm SNR_{G}}=5$ [see Eq.~(\ref{eq:SNR_G}) for definition]. A couple of notable trends is then summarized as follows:

\begin{itemize}
\item Overall, the impact of correlated magnetic noise, characterized by the ratio $|\langle S_{\rm B}\rangle_\epsilon|/\langle S_{\rm G}\rangle_{\rm SNR_G=5}$, decreases as the separation angle $\beta$ increases, but it is not monotonically changed.  
\item At some separations, we see a large variation of the ratio with respect to the parameter $\epsilon$, and the differences between the ratios depicted as green and blue or orange lines become eventually significant. 
\item The anisotropies in the magnetic noise spectrum, characterized by $\epsilon$, do not always worsen the impact of correlated magnetic noise, and the results in the isotropic case (green dashed lines) sometimes go below the blue or orange curves of $\epsilon=1$. 
\end{itemize}

These trends are exactly what we have seen in Sec.~\ref{subsec:impact_GWs}. Hence, in the presence of anisotropies in the magnetic field spectrum, the impact of correlated magnetic noise crucially depends not only on the geometric configuration, but also on the geographical locations for a pair of detectors.

Finally, we note that the VK pair tends to have a smaller value of the ratio, $|\langle S_{\rm B}\rangle_\epsilon|/\langle S_{\rm G}\rangle_{\rm SNR_G=5}$, than others, as shown in Fig.~\ref{fig:discussion}. Recalling the fact that the denominator of the ratio, i.e., $\langle S_{\rm G}\rangle_{\rm SNR_G=5}$, is recast as $5\sigma$ and it is basically determined by the convolution of the noise spectral density at each detector [see Eq.~(\ref{eq:variance}) with optimal filter given at Eq.~(\ref{eq:filter_function})], a smaller value of the ratio is partly ascribed to the low sensitivity of the GWs. In fact, in the original separation of the detector's pair, the cross-correlation amplitude $\langle S_{\rm G}\rangle_{\rm SNR_G=5}$ for the VK pair is  smaller than that for the HL pair by a factor of $13$. However, the numerator of the ratio, i.e., $\langle S_{\rm B}\rangle$, for the VK pair is found to be much smaller than that for the HL pair over various values of $\beta$, and its difference amounts to a factor of $700$ in the original setup. This means that for the VK pair, a large cancellation happens in the integral of $\langle S_{\rm B}\rangle$. Since the integrand is expressed as the product of the two oscillating functions, i.e., $M_{12}$ and $\widetilde{Q}$, we conclude that a large cancellation basically comes from multiple factors, including a detector's characteristic, geometric reasons, as well as the properties of the underlying GW signal. 

\section{Conclusion}
\label{sec:conclusion}

According to the event rate inferred from the currently detected GWs, we will be soon able to detect, via the second-generation detectors, many unresolved GW signals, viewed as a stochastic GW. Such an astrophysical origin, if detected, provides hints and clues for the formation and evolution of cosmological black hole or neutron star binaries. Increasing the sensitivity of laser interferometers, however, the correlated noise, detector's noise coupled with environmental disturbances that has a global correlation, is a potential concern, and can give a large impact on the detection of stochastic GWs. 

In this paper, we have presented a  comprehensive  study  of  the  impact of correlated noise at low-frequency bands, which appears through the coupling of the mirror control system with the stationary electromagnetic fields on the Earth, known as the Schumann resonances. In our previous work, we have proposed a simple analytical model that can characterize the impact of correlated magnetic noise on the detection of stochastic GWs. Albeit with several simplifications and assumptions, the model quantitatively described the key properties of correlated magnetic noise, which indeed match with those inferred from measurement by magnetometers. Then, we have explored the possible impact of the correlated noise on the ongoing and upcoming detectors. However, one important simplification that may possibly affect these estimates is the anisotropies in the lightning source distribution, or equivalently, the magnetic field spectrum. The present paper particularly considered this issue, and based on the observed lightning activity data, we examine if the anisotropies in the magnetic field spectrum can alter the previous estimates on the impact of correlated magnetic noise. 

Introducing a model of the magnetic field spectrum given at Eq.~(\ref{eq:P_B_anisotropies}) with (\ref{eq:weight-func}), we first see that the changes in the magnetic noise spectrum, $M_{12}$, defined at Eq.~(\ref{eq:M_12}), are mostly coherent and small, just in proportion to the parameter $\epsilon$ that controls the degree of anisotropies (Sec.~\ref{subsec:correlated_noise_spectrum}). However, the quantitative estimation of the correlated magnetic noise, quantified by the cross-correlation statistic $\langle S_{\rm B}\rangle$, reveals that the impact on the detection of stochastic GWs can change largely with $\epsilon$, depending on which type of GWs we observe (Sec.~\ref{subsec:impact_GWs}). As opposed to a naive expectation, the impact of correlated magnetic noise does not always increase with $\epsilon$, but it is rather suppressed to some extent for a specific detector's pair. One such case is the LIGO Hanford and Livingston pair. Also, we find that even in the presence of anisotropies, there is a robust detector pair for which the amplitude of correlated magnetic noise becomes comparable to or well below that of the detectable stochastic GWs, irrespective of the type of GWs. This is the Virgo and KAGRA (VK) pair. To better understand these results, we considered somewhat artificial situations, and found that in the presence of anisotropies, the properties of correlated magnetic noise crucially depend on both the geometrical and geographical setup of the detector's pair. For the VK pair, which could potentially achieve the best sensitivity to the stochastic GWs in the most pessimistic case, it is suggested that the detector’s characteristic, i.e., detector's intrinsic noise property is also another important factor to reduce the impact of correlated magnetic noise. 
To be more precise, the VK pair is less sensitive to the GW signals at $f\lesssim25$\,Hz, thus avoiding the low-frequency Schumann resonances that are the major concern to produce a large correlated noise. Rather, the VK pair is sensitive to a relatively high-frequency GW at $f\gtrsim40$\,Hz, where both the functions $M_{12}$ and $\gamma_{\rm 12}$ start to oscillate rapidly because of a largely separated pair. This helps further mitigate the correlated magnetic noise through the phase cancellation to the broadband noise contribution integrating over wide frequencies. These conditions may give a hint to build a third-generation detector robust against the Schumann resonances.

Finally, despite several nontrivial and interesting findings in this paper, we must admit that those results may  rely on the simplifications and assumptions made in our analytical model. From a conservative point of view, the results presented here have to be taken with caution. Nevertheless, the underlying reasons or explanation would be certainly relevant, and based on these, a more systematic calibration of the noise correlations has to be developed, and methodology to mitigate the correlated magnetic noise should be exploited.

\begin{acknowledgments}
This work was supported in part by MEXT/JSPS KAKENHI Grant Number JP18H04591 (Y.H.) and JP15H05899, JP16H03977, and JP17H06359 (A.T.). 
\end{acknowledgments}

\appendix

\section{Nulling condition for the coherence function 
$\gamma_\ell^{\rm B}$ }
\label{appendix:special}

In this appendix, we show that in the case of the magnetic field spectrum having a special anisotropy, there still exists the nulling condition for the coupling vectors $\widehat{\bm X}_i$ at each detector pair, under which the coherence function $\gamma_\ell^{\rm B}$ becomes vanishing, and hence the correlated magnetic noise is canceled. 

Let us first look for the nulling condition in the isotropic case (see also Ref.~\cite{2017PhRvD..96b2004H}). 
Substituting Eq.~(\ref{eq:unit_vect_ei}) into Eq.~(\ref{eq:gamma_integral}), we rewrite the coherence function $\gamma_\ell^{\rm B}$ with the following tensorial form:
\begin{align}
&\gamma_{\ell}^{\rm B}(\widehat{\bm r}_{1},\widehat{\bm r}_{2}) =
\frac{(2\ell+1)(\ell-1)!}{2\pi(\ell+1)!} \nonumber \\
&\qquad\qquad~\times 
\,\Gamma^{b e}(\ell, \widehat{\bm r}_{1}, \widehat{\bm r}_{2})\, \epsilon_{abc}\,
\epsilon_{def}\,{\widehat r}_{1}^{c}\,{\widehat r}_{2}^{f}\,{\widehat X}_{1}^{a}\,{\widehat X}_{2}^{d}\,,
\label{eq:gamma_tensor}
\end{align}
with $\epsilon_{abc}$ being the three-dimensional Levi-Civita symbol. Here, we adopt the Einstein's summation convention. The quantity $\Gamma^{ab}$ is the symmetric matrix defined by 
\begin{align}
\Gamma^{a b}(\ell, \widehat{\bm r}_{1}, \widehat{\bm r}_{2}) 
= \int_{S^{2}} \,d^2 \widehat{\bm \Omega}\, \frac{{\cal P}_{\ell}^{1}(\widehat{\bm \Omega} \cdot \widehat{\bm r}_{1})}{|\widehat{\bm \Omega} 
\times \widehat{\bm r}_{1}|}\,
\frac{{\cal P}_{\ell}^{1}(\widehat{\bm \Omega} \cdot \widehat{\bm r}_{2})}{|\widehat{\bm \Omega} 
\times \widehat{\bm r}_{2}|}\,{\widehat \Omega}^{a}\,{\widehat \Omega}^{b}\,.
\label{gamma_contract}
\end{align}
Because of its rotational covariance, the above matrix is, after integrating over the solid angle, expressed as a function of $\ell$ and the directional cosine between the position vectors $\widehat{\bm r}_{1}$ and $\widehat{\bm r}_{2}$, which we denote by $\mu\equiv\widehat{\bm r}_{1}\cdot\widehat{\bm r}_{2}$. We then express $\Gamma^{a b}$ as the most general tensor form constructed with $\delta_{ab}$ and $\widehat{r}_{a}$:
\begin{align}
 \Gamma^{a b}(\ell, \mu) &=  F_{\ell}(\mu) \delta^{ab} 
+ G_{\ell}(\mu)({\widehat r}_{1}^{a} {\widehat r}_{2}^{b}+{\widehat r}_{2}^{a} {\widehat r}_{1}^{b})
\nonumber \\
&\quad+ H_{\ell}(\mu)({\widehat r}_{1}^{a} {\widehat r}_{1}^{b}+{\widehat r}_{2}^{a} {\widehat r}_{2}^{b}).
\label{eq:gamma_sym}
\end{align}
The explicit form of the functions $F_\ell$, $G_\ell$, and $H_\ell$ are presented in Ref.~\cite{2017PhRvD..96b2004H}, but we do not need their functional forms to find the nulling condition. Rather, we substitute Eq.~(\ref{eq:gamma_sym}) into Eq.~(\ref{eq:gamma_tensor}), and rewrite the coherence function $\gamma_\ell^{\rm B}$ with
\begin{align}
&\gamma_{\ell}^{\rm B}(\widehat{\bm r}_{1},\widehat{\bm r}_{2}) = \frac{(2\ell+1)(\ell-1)!}{2\pi(\ell+1)!} 
\nonumber\\
&\qquad \times \Bigl[
F_\ell(\mu)\,\left\{\mu\,(\widehat{\bm X}_1 \cdot \widehat{\bm X}_2)
-(\widehat{\bm r}_{2} \cdot \widehat{\bm X}_1) \, (\widehat{\bm r}_{1} \cdot \widehat{\bm X}_2) \right\}
\nonumber \\
&\qquad
-G_\ell(\mu)\,\left\{(\widehat{\bm r}_1 \times \widehat{\bm r}_2) \cdot \widehat{\bm X}_1 \right\}
\, \left\{(\widehat{\bm r}_1 \times \widehat{\bm r}_2) \cdot \widehat{\bm X}_2 \right\}
\Bigr]\,.
\label{eq:gamma_ell_B_iso}
\end{align}
From this expression, the coherence function is shown to be zero if the projection vectors, $\widehat{\bm X}_i$ and $\widehat{\bm X}_j \, (i, j=1, 2 \,\, {\rm and} \,\, i\neq j)$,  satisfy the following relation (for any double sign):
\begin{align}
(\widehat{\bm X}_i\,,  \widehat{\bm X}_j) =
\left( \pm \frac{\widehat{\bm r}_i \times \widehat{\bm r}_j}{|\widehat{\bm r}_i \times \widehat{\bm r}_j|} \,, 
 \pm \frac{(\widehat{\bm r}_i \times \widehat{\bm r}_j)\times \widehat{\bm r}_j}
{|(\widehat{\bm r}_i \times \widehat{\bm r}_j) \times \widehat{\bm r}_j|} \right)\,.
\label{eq:nulling_vector}
\end{align}
Equation.~(\ref{eq:nulling_vector}) is the nulling condition that cancels the correlated magnetic noise, and it says that the coherence function vanishes when the two projection vectors $\widehat{\bm X}_i$ are orthogonal to each other, and one of them points to the direction parallel or perpendicular to the great circle connecting the pair of detectors.

Based on the nulling condition in the isotropic case, we next consider the anisotropic case, in which the coherence function is replaced with Eq.~(\ref{eq:gamma_integral2}), and the dependence of the magnetic field spectrum appears manifest. Assuming the functional form of Eq.~(\ref{eq:P_B_anisotropies}), the coherence function now depends on the function $W(\widehat{\bm \Omega})$.  Since this is the scalar quantity, any anisotropies in the magnetic field spectrum are expressed as the functions of $\widehat{\bm {\Omega}}\cdot\widehat{\bm{L}}_i$, with $\widehat{\bm{L}}_i$ being the projection vector. In principle, there must be multiple sets of projection vectors to express the general form of anisotropies. But, for illustrative purpose, we here consider the simplest case with the single projection vector, $\widehat{\bm{L}}$, so that the function $W$ is expressed in the form as $W(\widehat{\bm{\Omega}}\cdot\widehat{\bm{L}})$. This means that we impose the axial symmetry in the magnetic field spectrum.  Then, the symmetric matrix $\Gamma^{ab}$ given at Eq.~(\ref{gamma_contract}) is replaced with
\begin{align}
\Gamma^{a b}&(\ell, \widehat{\bm r}_{1}, \widehat{\bm r}_{2},\widehat{\bm{L}}) 
=\nonumber\\
&\int_{S^{2}} \,d^2 \, \widehat{\bm \Omega}\, W(\widehat{\bm{\Omega}}\cdot\widehat{\bm{L}}) \frac{{\cal P}_{\ell}^{1}(\widehat{\bm \Omega} \cdot \widehat{\bm r}_{1})}{|\widehat{\bm \Omega} 
\times \widehat{\bm r}_{1}|}\,
\frac{{\cal P}_{\ell}^{1}(\widehat{\bm \Omega} \cdot \widehat{\bm r}_{2})}{|\widehat{\bm \Omega} 
\times \widehat{\bm r}_{2}|}\,{\widehat \Omega}^{a}\,{\widehat \Omega}^{b}\,.
\label{gamma_contract2}
\end{align}
Making use of the same analogy as in the isotropic case, the rotational covariance suggests that the above matrix is, after performing the angular integral, expressed as the function of $\ell$, $\mu$, and $\nu_i=\widehat{\bm r}_i\cdot\widehat{\bm L}$\,\,$(i=1,2)$, and has the following tensorial form:
\begin{align}
 \Gamma^{a b}(\ell, \mu,\nu_1,\nu_2) &=  F_{\ell}\, \delta^{ab} 
+ G_{\ell}\,({\widehat r}_{1}^{a} {\widehat r}_{2}^{b}+{\widehat r}_{2}^{a} {\widehat r}_{1}^{b})
\nonumber \\
&\quad+ H_{\ell}({\widehat r}_{1}^{a} {\widehat r}_{1}^{b}+{\widehat r}_{2}^{a} {\widehat r}_{2}^{b})
\nonumber \\
&\quad+ I_{\ell}\,{\widehat L}^{a} {\widehat L}^{b}
+J_{\ell}\,({\widehat L}^{a} {\widehat r}_{1}^{b}+{\widehat r}_{1}^{a} {\widehat L}^{b})
\nonumber \\
&\quad+K_{\ell}\,({\widehat L}^{a} {\widehat r}_{2}^{b}+{\widehat r}_{2}^{a} {\widehat L}^{b})
\label{eq:gamma_sym2}
\end{align}
Note that all the scalar functions in the above (e.g., $F_\ell$, $G_\ell$, $H_\ell$, $\cdots$) depend not only on $\ell$, but also on $\mu$, $\nu_1$, and $\nu_2$. Substituting Eq.~(\ref{eq:gamma_sym2}) into Eq.~(\ref{eq:gamma_tensor}), the coherence function is now given in the following form:
\begin{align}
&\gamma_{\ell}^{\rm B}(\widehat{\bm r}_{1},\widehat{\bm r}_{2}, \widehat{\bm L}) = \frac{(2\ell+1)(\ell-1)!}{2\pi(\ell+1)!} 
\nonumber\\
&\qquad \times \Bigl[
F_\ell\,\left\{\mu\,(\widehat{\bm X}_1 \cdot \widehat{\bm X}_2)
-(\widehat{\bm r}_{2} \cdot \widehat{\bm X}_1) \, (\widehat{\bm r}_{1} \cdot \widehat{\bm X}_2) \right\}
\nonumber \\
&\qquad
-G_\ell\,\left\{(\widehat{\bm r}_1 \times \widehat{\bm r}_2) \cdot \widehat{\bm X}_1 \right\}
\, \left\{(\widehat{\bm r}_1 \times \widehat{\bm r}_2) \cdot \widehat{\bm X}_2 \right\}
\nonumber \\
&\qquad
+I_\ell\,\left\{(\widehat{\bm L} \times \widehat{\bm r}_1) \cdot \widehat{\bm X}_1 \right\}
\, \left\{(\widehat{\bm L} \times \widehat{\bm r}_2) \cdot \widehat{\bm X}_2 \right\}
\nonumber \\
&\qquad
+J_\ell\,\left\{(\widehat{\bm L} \times \widehat{\bm r}_2) \cdot \widehat{\bm X}_2\right\}
\left\{(\widehat{\bm r}_1 \times \widehat{\bm r}_2) \cdot \widehat{\bm X}_1\right\}
\nonumber \\
&\qquad
+K_\ell\,\left\{(\widehat{\bm L} \times \widehat{\bm r}_1) \cdot \widehat{\bm X}_1\right\}
\left\{(\widehat{\bm r}_1 \times \widehat{\bm r}_2) \cdot \widehat{\bm X}_2\right\}
\Bigr]\,.
\label{eq:tensorform_aniso}
\end{align}
Notice that the first two terms in the bracket are the same tensorial form as we saw in the isotropic case [see Eq.~(\ref{eq:gamma_ell_B_iso})]. Thus, in order to cancel the correlated magnetic noise (equivalently to set $\gamma_\ell^{\rm B}$ to zero), the condition given at Eq.~(\ref{eq:nulling_vector}) still needs to be satisfied as the sufficient condition. On top of this, the last three terms in the bracket have to be nulled, leading to the following additional constraints [substituting Eq.~(\ref{eq:nulling_vector}) explicitly]:
\begin{align}
(\widehat{\bm L} \times \widehat{\bm r}_j)\cdot\bigl\{(\widehat{\bm r}_i\times\widehat{\bm r}_j)\times\widehat{\bm r}_j\bigr\}=0,\,\,(i,j=1,\,2\, \mbox{and}\,\,i\ne j),
\label{condition1}
\end{align}
which states that in addition to the condition for coupling vectors $\widehat{\bm{X}}_i$, the symmetric axis $\widehat{\bm L}$ has to be also restricted, and it should be described by the linear combination of the vectors $\widehat{\bm{r}}_1$ and $\widehat{\bm{r}}_2$. 
In other words, the symmetric axis  $\widehat{\bm L}$ must lie on the plane spanned by $\widehat{\bm r}_1$ and $\widehat{\bm r}_2$.

Note that the constraint given at Eq.~(\ref{condition1}) does not necessarily hold for $\widehat{\bm L}$ to satisfy $\gamma_\ell^{\rm B}$=0. In fact, one can show that the following choice is also possible, leading to $\gamma_\ell^{\rm B}=0$ under the condition at Eq.~(\ref{eq:nulling_vector}):  
\begin{align}
\widehat{\bm L} = \frac{\widehat{\bm r}_1 \times \widehat{\bm r}_2}{|\widehat{\bm r}_1 \times \widehat{\bm r}_2|}.
\label{condition2}
\end{align}
That is, the symmetric axis, $\widehat{\bm L}$, is now perpendicular to the plane spanned by $\widehat{\bm r}_1$ and $\widehat{\bm r}_2$. Setting $i=1$ and $j=2$ in Eq.~(\ref{eq:nulling_vector}) and substituting Eq.~(\ref{condition2}) into Eq.~(\ref{eq:tensorform_aniso}), only the term proportional to $J_\ell$ is found to be algebraically nonvanishing. The remaining term, however, is also shown to become vanishing  
due to the periodicity of the integrand of $J_\ell$ as follows. Under the conditions given at Eqs.~(\ref{eq:nulling_vector}) and (\ref{condition2}) (with $i=1$ and $j=2$), we compare between the one derived from Eq.~(\ref{eq:gamma_tensor}) with Eq.~(\ref{gamma_contract2}) and Eq.~(\ref{eq:tensorform_aniso}). We obtain 
\begin{align}
&J_{\ell}(\widehat{\bm r}_{1},\widehat{\bm r}_{2}, \widehat{\bm L}) =
 \frac{1}{|\widehat{\bm r}_1 \times \widehat{\bm r}_2| |(\widehat{\bm r}_1 \times \widehat{\bm r}_2)\times \widehat{\bm r}_2|^{2}} 
\nonumber\\
&\qquad \times \int_{S^{2}} \,d^2 \, \widehat{\bm \Omega}\, W(\widehat{\bm{\Omega}}\cdot\widehat{\bm{L}}) \frac{{\cal P}_{\ell}^{1}(\widehat{\bm \Omega} \cdot \widehat{\bm r}_{1})}{|\widehat{\bm \Omega} 
\times \widehat{\bm r}_{1}|}\,
\frac{{\cal P}_{\ell}^{1}(\widehat{\bm \Omega} \cdot \widehat{\bm r}_{2})}{|\widehat{\bm \Omega} 
\times \widehat{\bm r}_{2}|}
\nonumber\\
&\qquad \times 
\Bigl[ \widehat{\bm{\Omega}}\cdot \left( \widehat{\bm r}_{2} -\mu \widehat{\bm r}_{1} \right) \Bigr]
\Bigl[ \widehat{\bm{\Omega}}\cdot \left( \widehat{\bm r}_{1} \times \widehat{\bm r}_{2} \right) \Bigr]\,.
\label{eq:j1}
\end{align}
The integrand of Eq.~(\ref{eq:j1}) is the periodic function of $\widehat{\bm \Omega}$ with respect to the rotation around the symmetric axis $\widehat{\bm L}$, and periodically changes its sign. To see this more clearly, without loss of generality, we introduce the following coordinate system for $\widehat{\bm \Omega}$:
\begin{align}
 \widehat{\bm \Omega} = \cos\theta\, \widehat{\bm{L}} +\sin\theta \Bigl( \cos\phi \, \widehat{\bm r}_{1} 
 + \sin\phi \, \widehat{\bm r}_1'\Bigr),
\label{eq:coordinate}
\end{align}
where $\theta$ and $\phi$ are the polar and azimuthal angles, respectively. The vector $\widehat{\bm r}_1'$ is the unit vector perpendicular to $\widehat{\bm r}_1$ and $\widehat{\bm L}$, and it is expressed as  $\widehat{\bm r}_1'=\left( \widehat{\bm r}_{2} -\mu \widehat{\bm r}_{1} \right)/\sqrt{1-\mu^{2}}$. Then, Eq.~(\ref{eq:j1}) is rewritten in the following form: 
\begin{align}
&J_{\ell}(\widehat{\bm r}_{1},\widehat{\bm r}_{2}, \widehat{\bm L}) =
 \frac{|\widehat{\bm r}_1 \times \widehat{\bm r}_2| }{|(\widehat{\bm r}_1 \times \widehat{\bm r}_2)\times \widehat{\bm r}_2|^{2}} 
\nonumber\\
& \times \int_{0}^{\pi} d \theta \, W(\cos \theta) \sin^{2}\theta \, \cos\theta \int_{0}^{2\pi} d \phi \, {\cal F}(\theta , \phi) \,,
\label{eq:j2}
\end{align}
with the function ${\cal F}(\theta , \phi)$ given by
\begin{align}
&{\cal F}(\theta , \phi) = \nonumber\\
& \sin\phi \frac{{\cal P}_{\ell}^{1}(\sin\theta \cos\phi)}{\displaystyle \sqrt{\cos^{2}\theta + \sin^{2}\theta \sin^{2}\phi }}
\frac{{\cal P}_{\ell}^{1}(\sin\theta \sin (\phi+\alpha))}{\sqrt{\cos^{2}\theta + \sin^{2}\theta \cos^{2}  (\phi+\alpha) }}.
\label{eq:j3}
\end{align}
Here, the angle $\alpha$ is defined through the relation, $\tan\alpha = \mu/\sqrt{1-\mu^{2}}$.
Using the parity symmetry of the associated Legendre polynomials [i.e.,  ${\cal P}_{\ell}^{1}(-x)=(-1)^{\ell+1}{\cal P}_{\ell}^{1}(x)$], 
the function  ${\cal F}$ is shown to have the following property, ${\cal F}(\theta , \phi+\pi)=-{\cal F}(\theta , \phi)$. Thus, the integral of the function ${\cal F}$ over the azimuthal angle vanishes, and hence the function $J_\ell$ becomes zero.

\begin{figure}[tb]
\begin{center}
\includegraphics[width=8.cm,angle=0,clip]{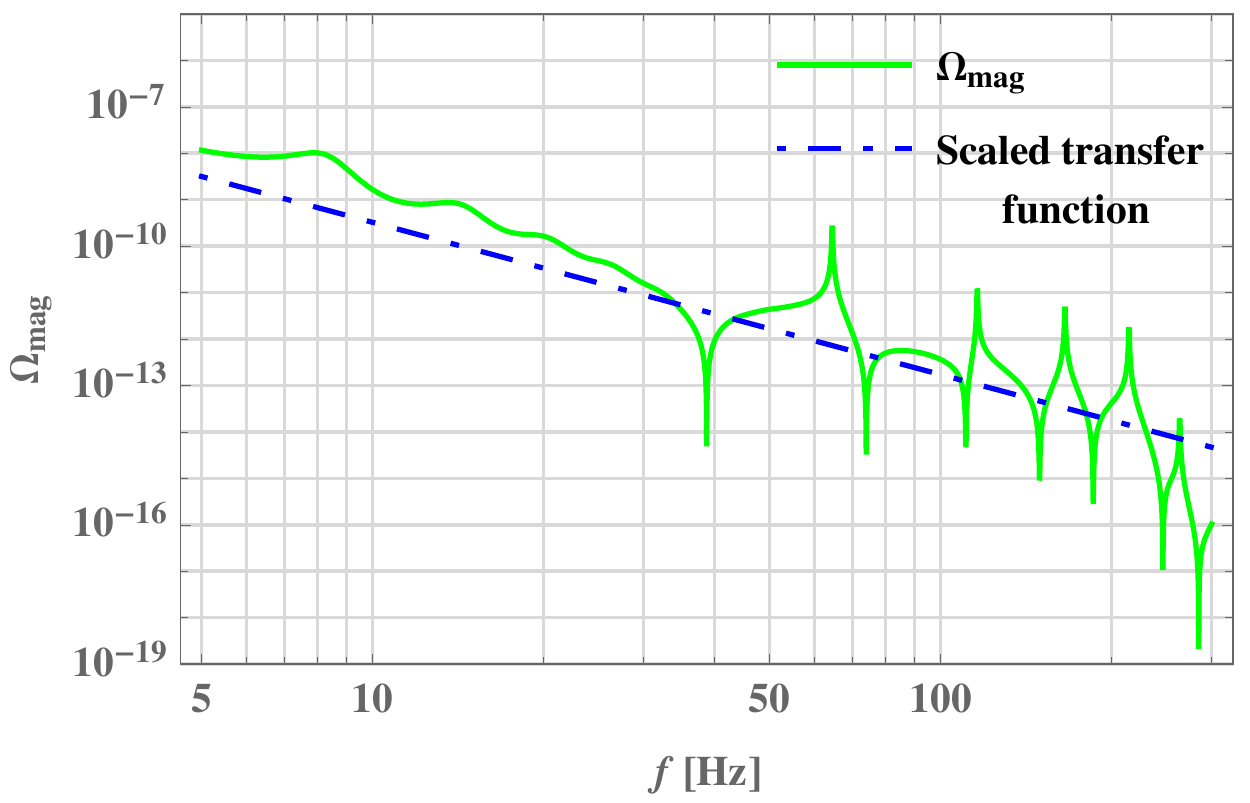}
\end{center}
\caption{Correlated magnetic noise in terms of an effective GW energy density $\Omega_{\rm mag}$ for the LIGO Hanford and Livingston pair (green solid). The function $\Omega_{\rm mag}$ is defined at Eq.~(\ref{eq:omegamag}) with (\ref{eq:gwdensity}). For comparison, the scaled transfer function, which corresponds to the blue dot-dashed line in Fig.~4 of Ref.~\cite{2019arXiv190302886T}, is also plotted (blue dot-dashed). The predicted behavior of $\Omega_{\rm mag}$ is similar to the measurement-based estimation shown in Ref.~\cite{2019arXiv190302886T} (depicted as filled points).  
}
\label{fig:omegamag}
\end{figure}

\begin{figure*}[tb]
\begin{center}
\hspace{-1cm}
\includegraphics[width=8.cm,angle=0,clip]{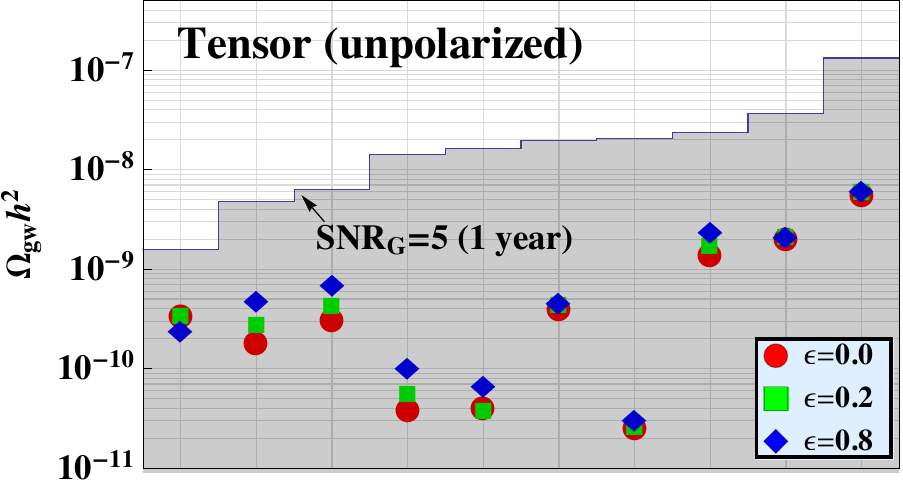}
\hspace*{5mm}
\includegraphics[width=7.4cm,angle=0,clip]{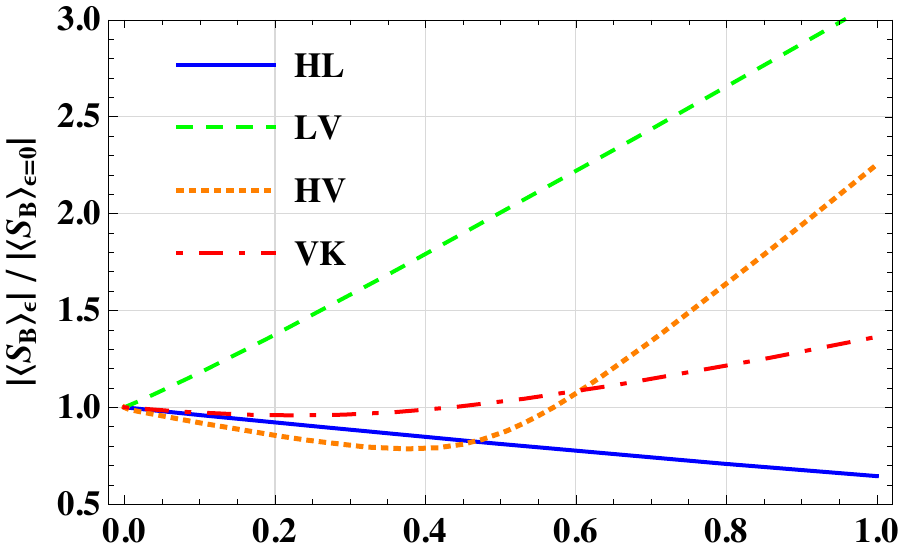}\\
\vspace*{0.0cm}

\hspace*{-29mm} 
{\bf \small HL \hspace{-0.1mm} HI \hspace{-0.1mm} LI  \hspace{-0.1mm} LV  \hspace{-0.4mm}  HV  \hspace{-1mm} VI  \hspace{-0.4mm}  VK   \hspace{-1mm} KI  \hspace{-1mm} HK  \hspace{-1mm}  LK } \hspace*{4.5cm}  $\epsilon$ \\
\end{center}
\caption{Same as in the upper panels of Fig.~\ref{fig:impact_flat_TITV}, but the results with the updated transfer function given by
Eq.~(\ref{eq:coupling})  with $(\kappa_i, b_i) = (0.079, 3.28)$. Note that the plotting range has been changed according to the resultant amplitudes of $\Omega_{\rm gw}h^2$.
}

\label{fig:impact_newcoupling}
\end{figure*}

\section{Correlated magnetic noise based on an updated transfer function of LIGO detectors}
\label{appendix:newcoupling}
In Sec.~\ref{sec:realistic}, setting the transfer function at    Eq.~(\ref{eq:coupling}) with $(\kappa_i, b_i)=(2, 2.67)$, we have estimated the impact of correlated magnetic noise. In this appendix, focusing on the unpolarized tensor mode and flat spectrum stochastic GWs, we adopt the recently calibrated transfer function of LIGO detectors in Ref.~\cite{Nguyen:2017}, and present the estimated results of the impact of correlated magnetic noise. Assuming the single power-law form, we obtained an approximate fitting function to the dataset of Ref.~\cite{Nguyen:2017}, which is reduced to Eq.~(\ref{eq:coupling}) with $(\kappa_i, b_i)=(0.079, 3.28)$. We assume that all the detectors have the same amplitude of transfer functions as obtained in the LIGO detectors.

In order to check if our choice of $(\kappa_i,b_i)$ gives a reasonable result, Fig.~\ref{fig:omegamag} plots the analytic prediction of frequency-dependent correlated magnetic noise $\Omega_{\rm mag}$ for the LIGO Hanford and Livingston pair (solid green), which is compared with Fig.~4 of Ref.~\cite{2019arXiv190302886T}. Here, the quantity $\Omega_{\rm mag}$ represents the effective GW energy density defined by [see also Eq.~(10) in Ref.~\cite{2019arXiv190302886T}]:
\begin{align}
\Omega_{\rm mag}(f)=\left| \frac{r_{1}(f)r_{2}(f) M_{12}(f)}{\gamma_{12}(f) S_{0}(f)} \right|,
\label{eq:omegamag}
\end{align}
where the function $S_{0}$ is given by
\begin{align}
S_{0}(f)= \frac{3H_{0}^{2}}{10\pi^{2}f^{3}}\,,
\label{eq:gwdensity}
\end{align}
with the present Hubble parameter $H_{0}$ set to $67.9\,$km\,s$^{-1}$\,Mpc$^{-1}$. Note that $\gamma_{12}$ is the overlap reduction function for the unpolarized tensor mode. In plotting the result, the orientation angles characterizing the coupling vector, $\widehat{\bm{X}}_i$, have been chosen so as to reasonably reproduce frequency-dependent features in $\Omega_{\rm mag}$ measured during the Advanced LIGO second observing run (i.e., Fig.~4 of Ref.~\cite{2019arXiv190302886T})\footnote{To be precise, the coupling vector, which is defined on the tangent plane at each detector,  was chosen such that it points to $153^{\circ}$ for Hanford, and $54^{\circ}$ for Livingston from the local east direction counterclockwise.}. For ease of comparison, the scaled transfer function is also shown with a blue dot-dashed line, which corresponds to the line shown in  Ref.~\cite{2019arXiv190302886T}. The predicted behavior of $\Omega_{\rm mag}$ explain quantitatively the trends shown in Fig.~4 of Ref.~\cite{2019arXiv190302886T}, and we see several low-frequency peaks of Schumann resonance as well as zero-crossing points at high-frequency bands, which basically come from the combination of $M_{12}$ and $\gamma_{12}$. 

Having confirmed that our model with the updated transfer function works to describe the major trend of $\Omega_{\rm mag}$ well, 
we next estimate the impact of correlated magnetic noise in a similar manner to Fig.~\ref{fig:impact_flat_TITV}. The estimated results of the impact are then translated into the amplitude of stochastic GWs, $\Omega_{\rm gw}h^2$, and are shown in Fig.~\ref{fig:impact_newcoupling}. Here, the projection vector for each pair of detectors are adjusted such that the impact of correlated magnetic noise is taken to be maximum. 

Overall, qualitative trends are similar to what has been seen in Fig.~\ref{fig:impact_flat_TITV} (upper left), except for the amplitude of $\Omega_{\rm gw}h^2$, which is now suppressed by a factor of
400 over all pairs of detectors. Other points to note may be that the impact of correlated noise becomes rather sensitive to the anisotropies at the LIGO Livingston and Virgo (LV) pair, while the LIGO Hanford and Virgo (HV) pair becomes less sensitive. In any case, the results suggest that the correlated magnetic noise is not a serious issue if all the detectors can achieve the same level of transfer function as seen in LIGO Hanford and Livingston.

\section{Impact of correlated magnetic noise on the detection of astrophysical GW backgrounds}
\label{appendix:astro}

\begin{figure*}[tb]
\begin{center}
\hspace{-1cm}
\includegraphics[width=8.cm,angle=0,clip]{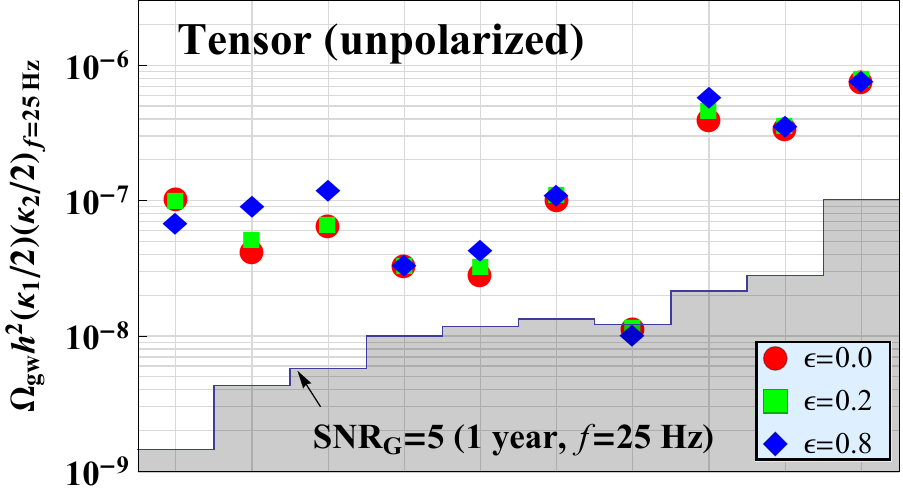}
\hspace*{5mm}
\includegraphics[width=7.4cm,angle=0,clip]{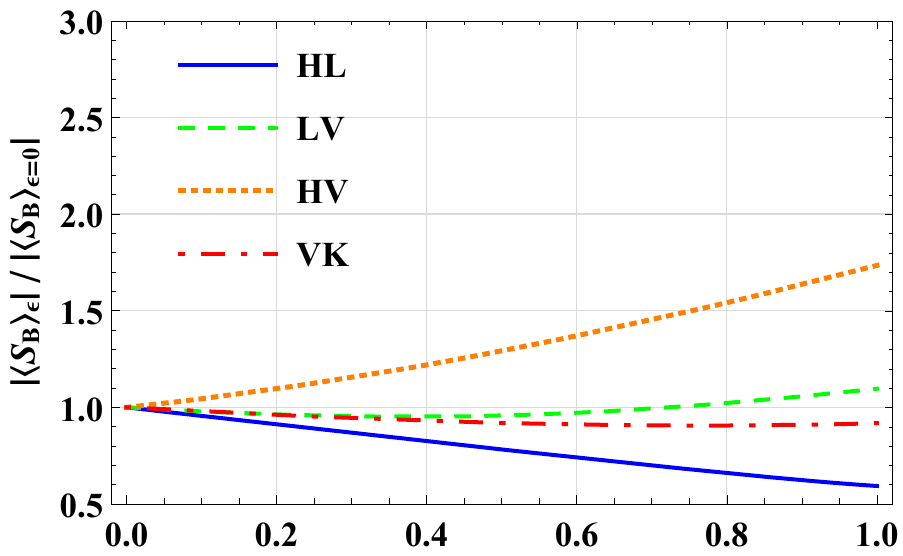}\\
\vspace*{0.0cm}
\hspace*{-29mm} 
{\bf \small HL \hspace{-0.1mm} HI \hspace{-0.1mm} LI  \hspace{-0.1mm} LV  \hspace{-0.4mm}  HV  \hspace{-1mm} VI  \hspace{-0.4mm}  VK   \hspace{-1mm} KI  \hspace{-1mm} HK  \hspace{-1mm}  LK } \hspace*{4.5cm}  $\epsilon$ \\
\vspace*{0.1cm}
\hspace*{-1cm} 
\includegraphics[width=8.cm,angle=0,clip]{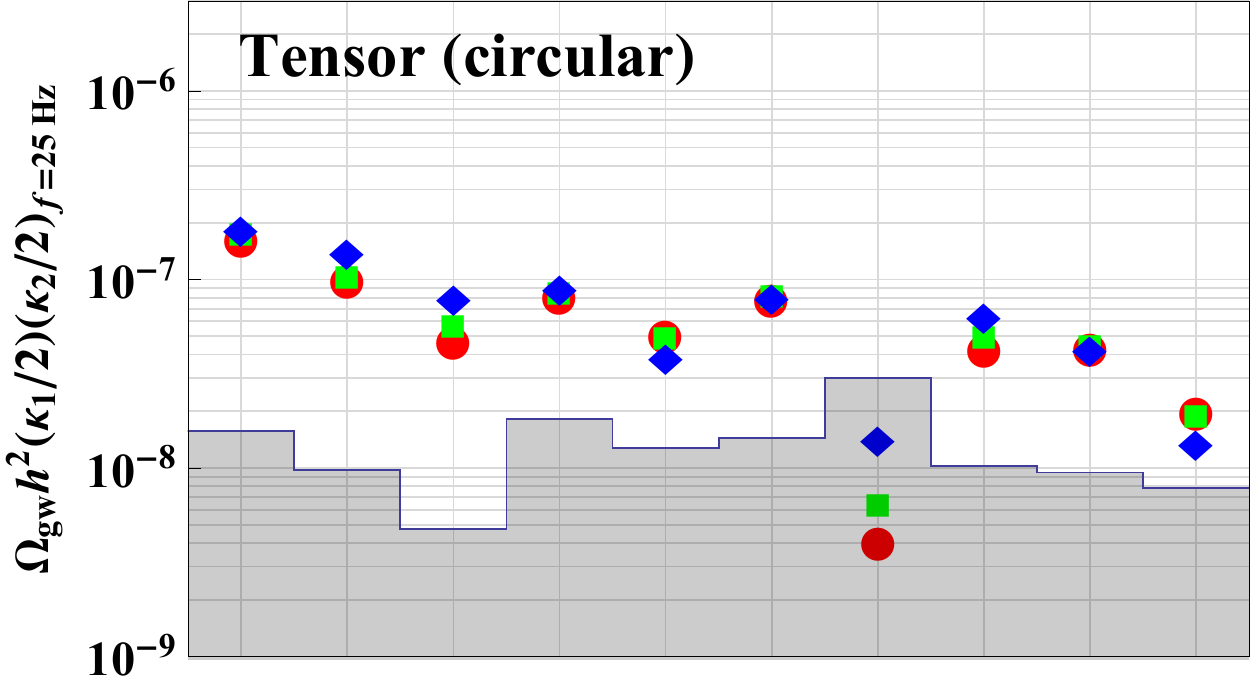}
\hspace*{5mm}
\includegraphics[width=7.4cm,angle=0,clip]{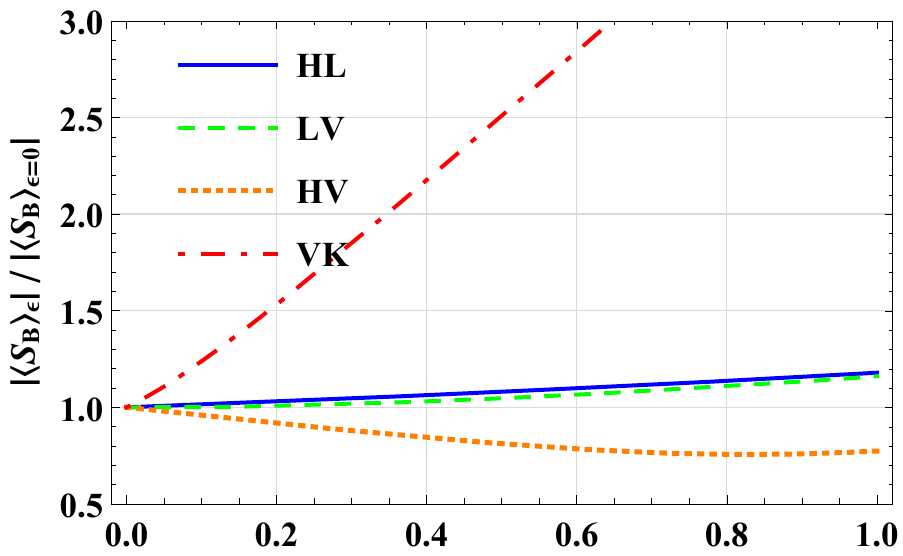}\\
\hspace*{8.5cm} $\epsilon$
\end{center}
\caption{Same as in Fig.~\ref{fig:impact_flat_TITV}, but here the impact of correlated magnetic noise on the detection of astrophysical GW backgrounds is shown, assuming the energy density spectrum of $\Omega_{\rm gw}^A\propto f^{2/3}$ with the pivot frequency  $f_0=25$\,Hz, i.e., $\Omega_{\rm gw}(f)=\Omega_{\rm gw,0}(f/f_0)^{2/3}$. Then, in the left panel, the estimated impact of the correlated magnetic noise is plotted as the amplitude at pivot frequency, $\Omega_{\rm gw,0}$. 
}
\label{fig:impact_astro_TITV}
\end{figure*}

\begin{figure*}[tb]
\begin{center}
\hspace{-1cm}
\includegraphics[width=8.cm,angle=0,clip]{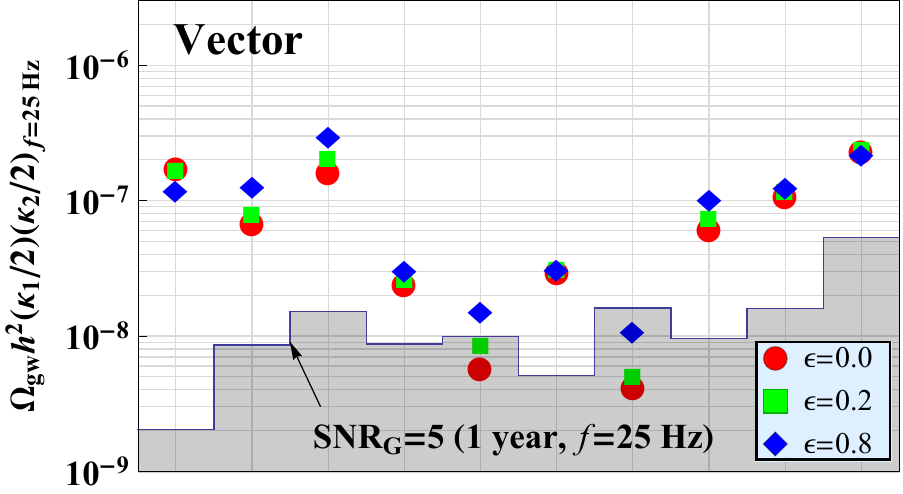}
\hspace*{5mm}
\includegraphics[width=7.4cm,angle=0,clip]{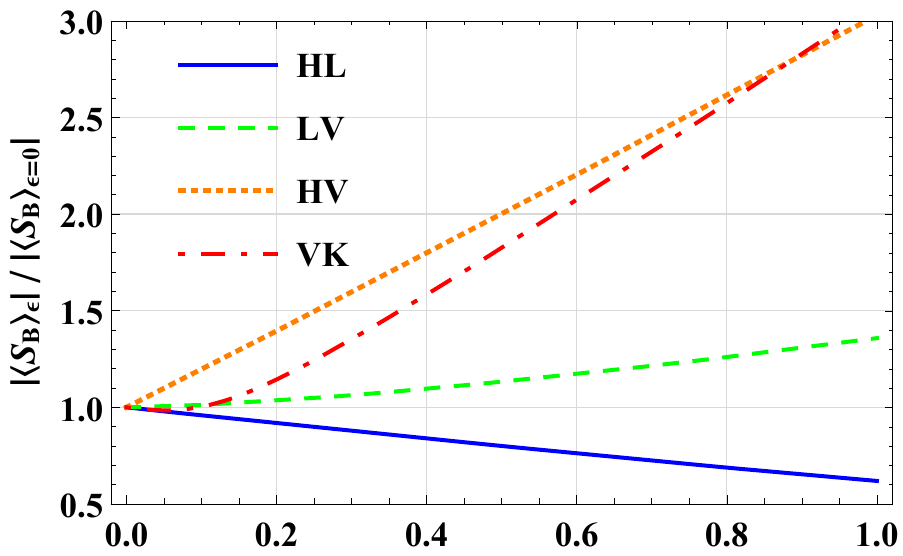}\\
\vspace*{0.0cm}
\hspace*{-29mm} 
{\bf \small HL \hspace{-0.1mm} HI \hspace{-0.1mm} LI  \hspace{-0.1mm} LV  \hspace{-0.4mm}  HV  \hspace{-1mm} VI  \hspace{-0.4mm}  VK   \hspace{-1mm} KI  \hspace{-1mm} HK  \hspace{-1mm}  LK } \hspace*{4.5cm}  $\epsilon$ \\
\vspace*{0.1cm}
\hspace*{-1cm} 
\includegraphics[width=8.cm,angle=0,clip]{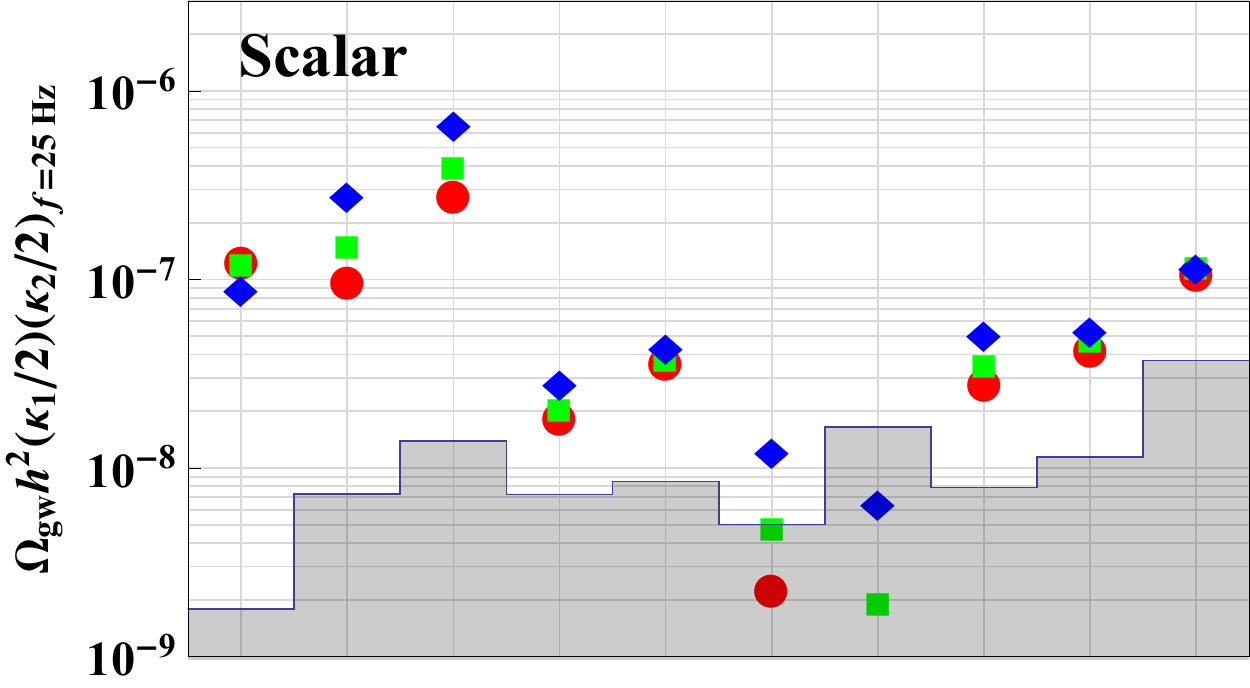}
\hspace*{5mm}
\includegraphics[width=7.4cm,angle=0,clip]{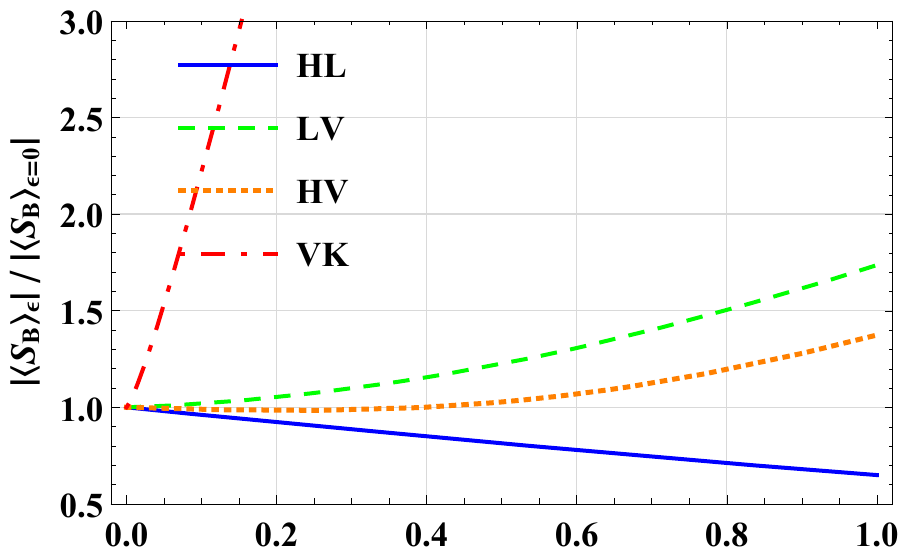}\\
\hspace*{8.5cm} $\epsilon$
\end{center}
\caption{Same as in Fig.~\ref{fig:impact_flat_SV}, but here the impact of the correlated magnetic noise on the detection of astrophysical GW backgrounds is shown, assuming the energy density spectrum of $\Omega_{\rm gw}^A\propto f^{2/3}$ with the pivot frequency  $f_0=25$\,Hz, i.e., $\Omega_{\rm gw}(f)=\Omega_{\rm gw,0}(f/f_0)^{2/3}$. Then, in the left panel, the estimated impact of the correlated magnetic noise is plotted as the amplitude at pivot frequency, $\Omega_{\rm gw,0}$. Note that in the lower left panel, the estimated value of $\Omega_{\rm gw,0}$ for the VK pair in the $\epsilon=0$ case is below the plotting region. } 
\label{fig:impact_astro_SV}
\end{figure*}

In Sec.~\ref{sec:realistic}, we have estimated the impact of correlated magnetic noise assuming the stochastic GW signals with flat spectrum, i.e., $\Omega_{\rm gw}\,\propto \,f^0$. In this appendix, we consider the stochastic GWs originated from the astrophysical sources having the power-law spectrum, $\Omega_{\rm gw}\propto f^{2/3}$, and present the estimated results of the impact of correlated magnetic noise. 

Similarly to Figs.~\ref{fig:impact_flat_TITV} and \ref{fig:impact_flat_SV}, the estimated impact of the correlated noise, $|\langle S_{\rm B}\rangle|$, is translated to the amplitude of stochastic GWs, $\Omega_{\rm gw}h^2$ at the pivot frequency $f=25$\,Hz, and is shown in Figs.~\ref{fig:impact_astro_TITV} and  \ref{fig:impact_astro_SV}.  Note that this frequency corresponds to the most sensitive band to detect stochastic GWs according to the design sensitivity of LIGO detectors.

Overall, qualitative trends are similar to what have been seen in Figs.~\ref{fig:impact_flat_TITV} and \ref{fig:impact_flat_SV}, but the estimated values of $\Omega_{\rm gw}h^2$ are found to be somewhat smaller. This is basically because the signal of the stochastic GWs now comes from the higher frequency range. On the other hand, the correlated magnetic noise characterized by the magnetic noise spectrum $M_{12}$ or coherence function $\gamma_\ell^{\rm B}$ remains unchanged, and it gives a large contribution at the low-frequency band. Hence, the  
optimal filter function involving the underlying spectrum of stochastic GWs [see Eq.~(\ref{eq:filter_function})] tends to pick up the higher-frequency band, and the contribution of the correlated magnetic noise is suppressed to some extent. Apart from these global trends, a couple of notable points are listed below:

\begin{itemize}
    \item For scalar-type GWs, the VI pair is less sensitive to the correlated magnetic noise, and possibly achieves the best sensitivity to the stochastic GWs, if the magnetic field spectrum is isotropic. 
    \item In contrast to the cases with underlying stochastic GWs having a flat spectrum, the correlated magnetic noise in the VK pair is rather sensitively affected by the anisotropies in the magnetic field spectrum, except for tensor-type unpolarized GWs. Nevertheless, the impact of correlated magnetic noise quantified by $\Omega_{\rm gw}h^2$ is well below the detectable amplitude of stochastic GWs with the signal-to-noise ratio of $5$ for the one-year observation (boundary between shaded and nonshaded regions), thus suggesting that the VK pair is robust against correlated magnetic noise irrespective of the underlying GW signals.    
\end{itemize}

\bibliographystyle{apsrev4-1}
\input{ms_Schumann_resubmit.bbl}

\end{document}

%% file: ms_Schumann_resubmit.bbl
%